\documentclass[12pt,preprint]{aastex}

\usepackage{CJK}
\usepackage{epsfig}

\shorttitle{Fulleranes in Circumstellar Environments}
\shortauthors{Zhang et al.}

\begin{document}

\title{Search for Hydrogenated C$_{60}$ (fulleranes) in Circumstellar Envelopes}

\begin{CJK*}{UTF8}{gbsn}

\author{Yong Zhang (张泳)\altaffilmark{1,2}, SeyedAbdolreza Sadjadi\altaffilmark{1}, Chih-Hao Hsia (夏志浩)\altaffilmark{4,1,2}, Sun Kwok (郭新)\altaffilmark{1, 3, 5}}

\affil{$^1$ Laboratory for Space Research, Faculty of Science, The University of Hong Kong, Pokfulam Road, Hong Kong, China\\
$^2$ Department of Physics, The University of Hong Kong, Pokfulam Road, Hong Kong, China\\
$^3$ Department of Earth Sciences, The University of Hong Kong, Pokfulam Road, Hong Kong, China\\
$^4$ Space Science Institute, Macau University of Science and Technology, Avenida Wai Long, Taipa, Macau, China\\
$^5$ Visiting Professor, Department of Physics and Astronomy, University of British Columbia, Vancouver, B.C., Canada
}
 \email{zhangy96@hku.hk; ssadjadi@hku.hk; chhsia@must.edu.mo; sunkwok@hku.hk}

\begin{abstract}

The recent detection of fullerene (C$_{60}$) in space and the positive assignment of five diffuse interstellar bands to C$_{60}^+$ reinforce the notion that fullerene-related compounds can be efficiently formed in circumstellar envelopes and be present in significant quantities in the interstellar medium. 
Experimental studies have shown that  C$_{60}$ can be readily hydrogenated, raising the possibility that hydrogenated fullerenes (or fulleranes, C$_{60}$H$_m$, $m=1-60$) may be abundant in space. 
In this paper, we present  theoretical studies of the vibrational modes of isomers of C$_{60}$H$_m$.  Our results show that the four mid-infrared bands from the C$_{60}$ skeletal vibrations remain prominent in slightly hydrogenated C$_{60}$, but their strengths diminish in different degrees with increasing hydrogenation. 
It is therefore possible that the observed infrared bands assigned to C$_{60}$ could be due to a mixture of fullerenes and fulleranes. This provides a potential explanation for the observed scatter of the C$_{60}$ band ratios.
Our calculations suggest that  a feature around 15\,$\mu$m due to the breathing mode of heavily hydrogenated C$_{60}$ may be detectable astronomically.  A preliminary search for this feature in 35 C$_{60}$ sources is reported.

\end{abstract}

\keywords{astrochemistry --- ISM: lines and bands --- infrared: stars --- stars: circumstellar matter}

\maketitle
\end{CJK*}

\section{Introduction}

It is now well established that circumstellar envelopes are major sources of molecules and solids in the Galaxy.  Recent infrared and millimeter-wave observations have shown that complex molecules can be efficiently synthesized in circumstellar envelopes of evolved stars, ejected into the interstellar medium, and distributed throughout the Galaxy \citep{kwok2004}.  The discovery of pre-solar grains carrying isotopic signatures of asymptotic giant branch stars offers proof that products of circumstellar chemical synthesis did survive journey across the Galaxy and be embedded in the early solar system \citep{zinner1998}.  
Among the molecules synthesized in the circumstellar environment is the pure carbon molecule fullerene (C$_{60}$).
Circumstellar C$_{60}$ was first discovered through its four vibrational bands in the young planetary nebula Tc1 \citep{cami10}. Subsequently, the C$_{60}$ vibrational bands have been detected in a variety of circumstellar environments \citep{sel10,gar10,gar11,Garcia11,gielen11,zk11,rs12,ev12,ots13}, although
the excitation and formation mechanisms of circumstellar C$_{60}$ remain unclear \citep[e.g.][and references therein]{zk13}.
Specifically, C$_{60}$ has been detected in 11, 4, and 7 planetary nebulae in the Galaxy, the Large Magellanic Cloud (LMC), and Small Magellanic Cloud (SMC),
respectively, and the detection rate is highest among the SMC planetary nebulae, followed by LMC and Galactic planetary nebulae \citep{gar12,ots16}. 
Some of the C$_{60}$ sources also exhibit vibrational bands from C$_{60}^+$ \citep{ber13,zk13,str15}.

Since its first synthesis in the laboratory, C$_{60}$ or its derivatives have been speculated to be abundant in the interstellar medium and may serve as the 
carrier of the diffuse interstellar bands (DIBs) \citep{kroto}. However,  the search for electronic transitions of C$_{60}$ in space has not been successful \citep{snow89,herbig00}, suggesting that interstellar C$_{60}$ (if any) might have been converted into other molecular forms. Based on the laboratory spectra recorded in a neon matrix \citep{fj93}, \citet{f94} ascribed two strong DIBs at 9632 and 9577\,{\AA} to C$_{60}^+$.  This identification has remained uncertain due to a lack of gas-phase experimental data and the non-detection of other DIBs coinciding with weaker C$_{60}^+$ bands. 
Recently, \citet{c15} reported the cold gas-phase spectrum of C$_{60}^+$, which exhibits features remarkably consistent with the two DIBs not only for the wavelengths but also for the band widths and relative intensities \citep[see also][]{c16a}. The weaker C$_{60}^+$ bands at 9348.4, 9365.2, and 9427.8\,{\AA} were soon thereafter discovered in diffuse clouds \citep{wb15,c16b}. Thus far five DIBs have been convincingly assigned to C$_{60}^+$.

The same C$_{60}^+$ absorption bands at 9632 and 9577\,{\AA} have been detected with remarkably large equivalent widths towards  a  C$_{60}$-containing proto-planetary nebula IRAS\,$01005$+$7910$ by \citet{ig13}, suggesting   that these two bands are primarily of circumstellar origin. Two DIBs at 4428 and 5780\,{\AA} were found to be unusually strong in C$_{60}$-containing planetary nebulae \citep{dg15}. These results strengthen the notion that there is a relationship between some DIB carriers and the fullerene-related compounds synthesized in circumstellar envelopes.

DIBs are likely to originate from  molecules or radicals possessing strong oscillator strengths. Given the high stability of the carbon cage, C$_{60}$ is able to survive in rather harsh environments \citep[see][and references therein]{cs09}. 
The atomic and electronic structures of C$_{60}$  make it possible to form a great diversity of stable derivatives, including C$_{60}$H$_m$ (fulleranes),  C$_{60}$ adducts, X$@$C$_{60}$ (endofullerenes; X represents one or more atoms or molecules), heterofullerenes, buckyonions, and so on. 
This is consistent with the general property of DIBs as DIBs can be divided into different groups according to the correlations of intensities. 
The possibility of various fullerene compounds as the carries of DIBs was recently discussed by \citet{om16}.

In the laboratory, C$_{60}$ can be readily hydrogenated into  C$_{60}$H$_{36}$ by atomic hydrogen \citep[see e.g.,][]{ci09,ig12}. 
Since hydrogen is the most abundant element in the universe and fullerene has been discovered in H-rich circumstellar envelopes \citep{gar10}, it is reasonable to expect that C$_{60}$H$_m$  might exist in space. Most of the experimental synthesis procedures of C$_{60}$H$_m$ are 
performed on condensed samples \citep{gm97} and it is difficult to say whether the radiative association reaction of gas-phase C$_{60}$ and H can efficiently produce C$_{60}$H$_m$.  
However, chemistry in astronomical environments  can be quite different from that in laboratory and it is possible that circumstellar C$_{60}$ exists in either or both solid-state and gas-phase forms   \citep{zk13}. 
Therefore, we cannot rule out the possibility that in space C$_{60}$H$_m$ can be formed on solid surfaces. 

Experimental studies suggest that C$_{60}$H$_m$ would be quickly dehydrogenated to C$_{60}$ in an inert gas atmosphere when the temperature 
approaches 550\,K \citep{ru93}.  If astronomical fulleranes are free-flying gaseous species, temperatures well above 550\,K can be readily reached through stochastic heating by absorption of a single UV photon.   If this is the case, significantly hydrogenated fullerenes would presumably only be present in the environments which are rich in atomic H and devoid of UV radiation.  While these  two conditions have a relatively narrow intersection, they are satisfied by the circumstellar environment of  reflection nebulae and proto-planetary nebulae. 

Experimental data show that the C$-$H stretching bands of C$_{60}$H$_m$ at 3.4--3.6\,$\mu$m have molar extinction coefficients similar to those of four major vibrational  bands of C$_{60}$ \citep{ig12}. These features were tentatively detected in the proto-planetary nebula IRAS\,01005$+$7910, and the relative intensities suggest a large $m$  value \citep{zk13}. 
However, \citet{dg16}  did not detect fulleranes in two planetary nebulae exhibiting strong C$_{60}$ emission, and concluded that they might have been destroyed by UV radiation. Indeed, the experimental study of \citet{ci09} shows that molecular hydrogen can be released from C$_{60}$H$_m$ by UV radiation and the C$_{60}$H$_m$/C$_{60}$ abundance ratio depends on the equilibrium between the hydrogenation of C$_{60}$ and the emission of molecular hydrogen.

Since spectral information on fulleranes from laboratory data is limited, it would be useful to perform theoretical studies  to explore the vibrational signatures of C$_{60}$H$_m$.  Recent advances in computational chemistry methods and computing facilities enable theoretical investigation of the vibrational properties of large molecules like fulleranes.   In this paper, we present theoretical calculations of the infrared spectra of C$_{60}$H$_m$, which then are compared with observational data. 
Specifically, we  investigate the effects of hydrogenation on the spectral behavior of C$_{60}$ and whether the spectral signatures can be detected by astronomical observations.

\section{C$_{60}$H$_m$ Isomers and Computational Methods}
\label{method}

Under laboratory conditions, C$_{60}$H$_{36}$ is the most common product to be synthesized although other fulleranes can also be produced.  Thermal annealing can partly eliminate hydrogen from  C$_{60}$H$_{36}$ to form C$_{60}$H$_{18}$ \citep{ig12}.  
The laboratory methods  to synthesize fulleranes  are summarized by \citet{gm97} and \citet{ca10}. Experimentally, C$-$H bonds are found to form through the addition of hydrogen across C$=$C double bonds, resulting in the numbers of hydrogen atoms in neutral fulleranes being always even. Guided by these experimental results, we have chosen to theoretically study eleven types of C$_{60}$H$_m$, where $m=2,4$...,20 and 36. Larger fulleranes ($m>36$) are highly unstable due to significant structural strain.

Nevertheless, we cannot rule out the presence of fullerane radicals with odd hydrogen atoms in astronomical environments.  
{ In laboratory conditions where there are plenty of H atoms available, C$_{60}$ is hydrogenated in the condensed phase, which is not necessarily the case in space.}
It is not unambiguously clear whether the dehydrogenation of fulleranes is accompanied by the release of H atoms or H$_2$ molecules. The calculations of \citet{zb12} show that the migration hydrogen on fulleranes have high barriers, unless the presence of a neighbouring H atom
lowers the migration barrier. 
{ If this is the case, the desorption of H$_2$ molecules would be more likely. Assuming that the formation route mimics the lab experiment, in which the most highly hydrogenated stable fullerane C$_{60}$H$_{36}$ is formed from C$_{60}$ and then the lower hydrogenation species are created by losing H$_2$, one will have only fulleranes with an even number of attached H atoms. 
Considering the relative difficulty of studying fulleranes with an odd number of H atoms (see below), it is desirable to begin with fulleranes with an even number of H atoms. 
Nevertheless, it would be useful to include fulleranes with an odd number of H atoms in future studies of association reactions of C$_{60}$H$_m$ with H atoms, dissociation paths, and their rates under space conditions.}

\begin{figure*}
\epsfig{file=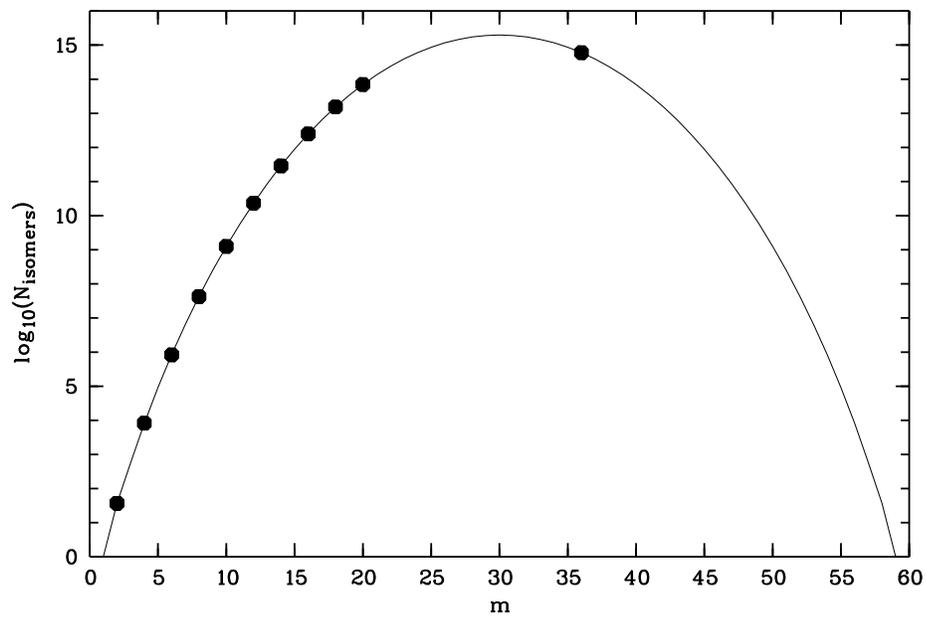, height=8cm}
\caption{Number of C$_{60}$H$_m$ isomers with increasing $m$ values.  The filled circles mark the fulleranes considered in this paper.
}
\label{isonum}
\end{figure*}

Mathematically, the number of fullerane isomers can be obtained using a combinatorial approach. In Figure~\ref{isonum}, we 
show the isomer numbers with increasing hydrogenation levels based on numbers in Table~2.5 of \citet{br95}.
As we can see from Figure \ref{isonum}, the number of structural isomers for each C$_{60}$H$_m$ is enormous ($N\sim10^{15}$ for $m=30$).  It would be computationally virtually impossible to explore all possible geometries.  We therefore limit ourselves to study only 5 isomers for each species (except for C$_{60}$H$_{18}$, for which 20 isomers are included).  While 5 is a very small fraction of the total isomer number, we note that the calculations of \citet{br95} do not take chemical stability into consideration.  In the present calculations, we focus on those isomers that are likely to be the most stable, and an investigation of the effect of limited isomer numbers 
will follow in Section~\ref{variation}.

The initial geometries of the isomers were generated based on the most stable isomers reported in the experimental and theoretical works of \citet{ca10}.
In general, we  start from C$_{60}$H$_{2}$ isomer and hydrogen atoms were added sequentially to generate the other fulleranes isomers. The initial structures were visualized using the Chemcraft graphical program\footnote{Available at http://www.chemcraftprog.com} to avoid similar geometries among isomers.
{ For highly hydrogenated C$_{60}$, H atoms are distributed as uniformly as possible over the C$_{60}$ cage to minimize the effect of the repulsive forces between H atoms.}

The computational procedures of C$_{60}$H$_m$ vibrational spectra are similar to those described in \citet{sz15}.  Density functional theory (DFT) calculations were conducted with the Gaussian 09  \citep{fr09}, PQS\footnote{Parallel Quantum Solutions version 4.0, 2013 Green Acres Road,  Fayetteville,  Arkansas  72703;  URL: http://www.pqs-chem.com} and Firefly\footnote{Firefly version 8.2.0,  http://classic.chem.msu.su/gran/firefly/index.html} packages, running on the HKU supercomputer grid-point facility and QS128-2300C-OA16 QuantumCubeTM machine separately.  We used the B3LYP  and BH\&HLYP hybrid functionals in combination with polarization consistent basis set PC1 \citep{je01,je02}  to obtain the harmonic frequencies of fundamental vibrations of the molecules.  
Under the default criteria, all the optimized geometries were characterized as local minima, established by the positive values of all harmonic frequencies and their associated eigenvalues of the second derivative matrix.  
Figure~\ref{structure} shows the optimized structures of C$_{60}$H$_m$ isomers included in this study.  The isomers are arranged from left to right in order of increasing relative total energy.

The scheme of double scaling factors \citep{lc12} was then applied to the DFT vibrational frequencies. In this scheme, the vibrational frequencies of $>1000$\,cm$^{-1}$ and $<1000$\,cm$^{-1}$ are scaled by 0.9311 and 0.9352 for BH\&HLYP and by 0.9654 and 0.9808 for B3LYP, respectively. 
In order to compare the computed absorption spectra with the observed emission spectra, we assumed a thermal excitation model at a temperature of 400\,K, and applied a Drude broadening function to the theoretical spectra of C$_{60}$H$_m$ isomers. 
{ The width is assumed to be about 0.3\,$\mu$m, in accordance with the observed widths of C$_{60}$ features.}
 The assumed temperature is comparable to those determined from astronomical spectra.  Finally, the theoretical spectrum of each C$_{60}$H$_m$ was obtained by averaging its five isomer spectra.

The accuracy of such combination of DFT functionals, basis set and scaling factors were estimated as 0.12--0.13\,$\mu$m in reproducing experimental wavelengths \citep{sz15}. The uncertainties of the band strengths are difficult
to estimate.  Inconsistent band strengths of C$_{60}$ have been reported in the literature, and there are large discrepancies between the theoretical and experimental values \citep[see][and references therein]{zk13,bg16}. Nevertheless,  DFT has been evaluated as having an excellent performance in predicting the relative infrared intensities and Raman activities \citep{Zvereva2011}.
From our calculations, we obtain the relative intrinsic strengths \footnote{{ In this paper, if not mentioning `intrinsic strengths', all the band intensities or intensity ratios refer to those modelled.}} of the C$_{60}$ bands at 18.9, 17.4, 8.5, and 7.0\,$\mu$m to be 100, 37, 31, and 71, lying well within the range of the previously reported values. 
The computed spectra of C$_{60}$ are compared with the Fourier-transform infrared data from Cataldo (priv. comm.) in Figure~\ref{com_c60}.  The reasonable agreement between theory and experiment gives us  confidence in the theoretical results.
 

Modeling the electronic ground states of the open-shell systems such as fullerane radical cations via DFT formalisms requires additional efforts. In order to eliminate the spin contamination from the wave function occurring in unrestricted DFT formalism such as UB3LYP, we set our model to the restricted open-shell  formalism. The second derivative energy matrix is calculated by numerical methods in the latter method rather than analytical methods in the former, which need more computational time. 
We tested the reliability of our chosen restricted open-shell approach in predicting the correct electronic structure by calculating the first ionization potential of C$_{60}$ from the energy difference of C$_{60}$ and C$_{60}^+$. The geometry optimization and frequency calculations on C$_{60}^+$ were performed with ROB3LYP/PC1. The calculated ionization energy is 7.85 eV, which is in excellent agreement with the experimental value of 7.65$\pm$0.20 eV reported in \citet{pog2004}.
Results of a limited study of cations are presented in Section~\ref{ions}.

\section{Results}

From Figure~\ref{structure} we can see that the geometry of the carbon cages is distorted by $sp^2$ to $sp^3$ hybridization change of carbon atoms undergo hydrogenation in different part of the C$_{60}$ cage.  
In the classical view of interacting hard-sphere atoms, the repulsion between H atoms is expected to increase the strain energy in cage, which may cause the instability and further decomposition of very heavily hydrogenated C$_{60}$.
As results of hydrogenation and rearrangement of $\pi$ bonds, different kinds of C$-$C bonds such as conjugated, olefinic and aromatic bonds are formed. The resultant breaking of the symmetry of C$_{60}$ cage leads to the activation of more C$-$C modes and changes the relative strengths of the four skeletal vibrational bands of C$_{60}$,  as well as forming new bands.
The addition of H atoms to the  C$_{60}$ molecule is also expected to introduce new C$-$H stretching and bending modes, with corresponding increase in intensities with the number of H atoms.   These effects will be quantitatively examined here.

\subsection{The C$-$H stretching mode at 3.4\,$\mu$m}

Figure~\ref{3um} shows the calculated profiles of the C$-$H stretching bands near 3.4\,$\mu$m. These stretching modes result in multiple peaks ranging from 3.3 to 3.9\,$\mu$m, and there is a general tendency of increasing band widths with increasing $m$ values. This may be attributed to the influence of neighboring environments of C$-$H groups. For a given C$-$H bond, the vibrational frequencies change depending on the number and distance of adjacent carbon atoms that are bonded to hydrogen \citep{web92}. 
When the $m$ value is small, increasing hydrogenation leads to more complexity, and thus more complex band profiles. Beyond a certain $m$ value, further hydrogenation results in less numbers of vacant sites and decreases the complexity. Therefore, the 3.4 $\mu$m band profile is a reflection of the symmetry of the molecule and can be taken as a probe of the structures of C$_{60}$H$_m$. 



The 3.4\,$\mu$m band arising from less symmetric C$_{60}$H$_m$  might be very broad and may be difficult to detect in astronomical spectra.  A mixture of C$_{60}$H$_m$ isomers may produce a broad plateau around 3.4\,$\mu$m, a feature that has been commonly observed in astronomical sources.

The strengths of the 3.4\,$\mu$m band are positively correlated with the degree of hydrogenation, as illustrated in Figure~\ref{3um}.   If C$_{60}$ is only slightly hydrogenated, the 3.4\,$\mu$m band may be too weak to be detectable in astronomical spectra.
In Figure~\ref{3um_m}, we examine the variations of the intensities with the $m$ values. A  positive correlation can be seen although the relation is clearly non-linear. When $m\le20$, the band intensities per hydrogen atom increase with increasing $m$ values.  
{ This shows that the C--H oscillators do not remain isolated from one another. The presence of additional H atoms affects the distribution of electrons, which respond to the distortion along C--H stretches and yield larger variations of the electric dipole moment.}

Experimental data suggest that the C$-$H stretching bands of C$_{60}$H$_{36}$ and C$_{60}$H$_{18}$ have molar extinction coefficients (normalized with the number of CH groups) larger  than those of aliphatic molecules \citep{ig12}. Therefore, if fullerenes with high hydrogenation have an abundance comparable to aliphatic species, they should be easily detected in astronomical spectra through the 3.4\,$\mu$m band.
Four bands around 3.4\,$\mu$m in the spectrum of IRAS\,01005$+$7910
have been tentatively assigned to fulleranes, and their relative intensities
might suggest that there exist species with different degrees of hydrogenation 
\citep{zk13}.  An inspection of Figure~\ref{3um} clearly suggests that it is possible to qualitatively reproduce the observed features by weighted sums of these computed spectra, while a quantitative comparison between theoretical and observational spectra requires more extensive computations of C$_{60}$H$_m$ isomers.


As C$_{60}$H$_m$ has a rather low ionization potential, C$_{60}$H$_{m}^+$ may be present in diffuse interstellar medium and contribute to the DIB phenomenon.  Theoretical computations and experimental data of polycyclic aromatic hydrocarbons (PAHs) show that the C--H stretching mode is greatly suppressed in cations \citep{pt95}.  Topological analysis of \citet{pp10} shows that a hole in the $\pi$ shell created by ionization may be responsible for the suppression of C--H modes in PAH cations.   It is therefore worthwhile to investigate whether such a suppression also occurs for fulleranes.  A detailed study of the vibrational spectrum of C$_{60}$H$_m^+$ is beyond the scope of this paper, but should be pursued in the future.    A preliminary study of two ionized species of fulleranes is given in Section \ref{ions}.

\subsection{C--H bending and carbon skeleton motions in the 5--24\,$\mu$m spectral region}

Due to the intrinsic weakness in C--H stretching intensities of the fulleranes with low hydrogen content, the non-detection of the 3.4\,$\mu$m in two C$_{60}$ sources \citep{dg16} cannot be used to infer their absence, and the search should be focused on the spectral range at longer wavelengths.   Figure~\ref{comp} shows the  theoretical spectra of C$_{60}$ and C$_{60}$H$_m$ in the 5--24\,$\mu$m region.  For comparison, we have also plotted the continuum-subtracted {\it Spitzer} spectrum of IRAS\,01005$+$7910 \citep[see][for details]{zk11}.  
Because of its high degree of symmetry, C$_{60}$ has only four infrared-active and ten Raman-active vibrational modes. All the infrared-active bands, except the one badly blended with the Unidentified Infrared Emission (UIE) band at 8.6\,$\mu$m, are clearly detected in IRAS\, 01005$+$7910. With hydrogenation, the symmetry of C$_{60}$ backbone is broken, and thus some Raman-active modes become infrared-active.  It is expected that  C$-$H bending coupled with C$_{60}$ skeletal vibrational modes could produce  a large number of bands in this spectral region.

From Figure~\ref{comp}, we can see broad emission features around  8, 15, and 20\,$\mu$m in most of the fulleranes,  which are qualitatively consistent with the experimental spectra of \citet{ig12}. Indeed, these broad features are also present in the spectrum of IRAS\,01005$+$7910.  However, since plateau features, especially those at 8 and 12 $\mu$m, can also be manifestation of aliphatic C$-$H bending modes \citep{kwok2001} { or the effects of anharmonicity and Fermi resonance in hot ordinary PAH \citep[see][for a study of the plateau in the 3\,$\mu$m region]{mp15}}, we cannot uniquely attribute the plateau features in IRAS\,01005$+$7910 to C$_{60}$H$_m$.
Below, we will discuss the origin of all three features.

The 8 $\mu$m feature:  vibrations around 8 $\mu$m are generally due to  C--H bending modes, along directions { parallel to} the tangential plane of the carbon sphere.  A broad 6--9\,$\mu$m feature is generally apparent in the spectra of fullerene-rich planetary nebulae in the Magellanic Clouds \citep{gar12}, which seems to resemble the theoretical spectra of fulleranes.

The 15\,$\mu$m feature: this feature arises from the breathing mode (sometimes referred to as radial mode) of carbon atoms where the skeleton inflate and deflate, coupled with peripheral C$-$H bending motion.  A schematic illustration of this mode for C$_{60}$H$_{36}$ is  shown in Figure~\ref{stru}.

The 20 $\mu$m feature:  vibrational modes in the 20 $\mu$m region are generally carbon skeleton deformation modes, where the angular and length  separations between carbon atoms change with vibration.  There is a well-known unidentified broad emission feature -- known in the literature as the 21-$\mu$m feature -- in this region   \citep{kwo89}.  
The feature has been attributed to fulleranes by \citet{web95} based on force-field and elastic-shell models.
Our calculations show that fulleranes can indeed produce broad features spanning the wavelength range 17--23\,$\mu$m. 
However, the profiles and peak positions of fulleranes vary greatly from one species to another, in contrast to the consistant peak wavelength and feature profiles of the observed  21\,$\mu$m feature  \citep{vol99}. 
We are unable to reproduce the observed 21-$\mu$m feature profiles with one, or the sum of more, theoretical fullerane spectra. 
This, together with the fact that there is no correlation between the 21\,$\mu$m feature and C$_{60}$ \citep{zk10,zk11}, lead us to conclude that fulleranes are not the carrier of the 21\,$\mu$m feature.

\subsection{Spectral variation among isomers of C$_{60}$H$_{18}$}\label{variation}

In order to assess the degree of spectral variation in different fullerane isomers, we have expanded the number of isomer from 5 to 20 for C$_{60}$H$_{18}$ (Figure \ref{c60h18geo}). The relative energy values of these twenty isomers expand a range of 80 kcal/mol (corrected for zero point energy).  Together with direct comparison of molecular geometries parameters,  this ensures the acceptable structural diversity of among these isomers.  The theoretical spectra of these 20 isomers are shown in Figure \ref{c60h18}.   Also plotted is the area (shown in grey) covered by one-standard-deviation of the 20 spectra.  All four dominant features around 3, 8, 15, and 20 $\mu$m are present, regardless of the variation in $m$ values.  These results suggest that the existence of the features is not specific to a particular isomer.
Specifically, the broad feature around 8 $\mu$m is the strongest one, followed by the feature that  peaks around 15 $\mu$m.
Unlike the 15 $\mu$m feature, the features around 20 $\mu$m vary greatly in intensities and peak positions among the isomers.
The intensities of the 3.4 $\mu$m feature are not sensitive to the specific isomer.


\subsection{Spectra of fullerane cations}\label{ions}

In order to test the possible effect of ionization on the spectral behavior of fulleranes, we have computed the spectra of two ionized species with low and high degree of hydrogenation : C$_{60}$H$_2^+$ and  C$_{60}$H$_{36}^+$  starting from the geometries of their lowest energy neutral isomers (isomer1). 
{ It is well known that due to theoretical limitations in dealing with the electronic energy states close to the ground one that arise in fullerenes and fulleranes when they have an open-shell electronic structure}, the  wavelengths and intensities of the C--C vibrations are not accurate \citep{pa11}. Consequently, we focus only on the C--H stretching bands in the 3--4\,$\mu$m wavelength range.  
From results of DFT calculations, we obtain co-added intensities of 8.4305, 21.0465, 1265.5103, and 533.8743\,km\,mol$^{-1}$ for the  3--4\,$\mu$m features of C$_{60}$H$_2$, C$_{60}$H$_2^+$, C$_{60}$H$_{36}$, and  C$_{60}$H$_{36}^+$, respectively. 
The theoretical spectra of these molecules are shown in Figure \ref{cations}. 
We can see from the figure that the peak positions and profiles appear to be invariant between neutral and ionized fulleranes, which could be attributed to the same geometries and symmetries. However, the change in intensity of the 3--4\,$\mu$m features upon ionization is not simple. 
C$_{60}$H$_{36}^+$ shows a decrease in intensity of its C--H stretching bands, while the opposite is true for C$_{60}$H$_2^+$. Such a spectral behavior is different from that of PAH molecules where the C--H stretching modes are suppressed in ions. 
Such suppression is not a direct result of ionization, but rather the result of  changes of electronic structures upon ionization  \citep{pp10}. Specifically, band intensities reflect the distortion of electronic density with internal displacements of the nuclei. 
More thorough studies of fullerane cations are required to determine the effects of ionization and how such changes depend on specific isomers and hydrogen contents.

\section{Discussions}

\subsection{Origin of the scatter of observed band ratios in C$_{60}$ sources}

The fullerene (C$_{60}$) molecule was detected through their four infrared-active bands due carbon skeletal modes.  These four bands  are still visible for slightly hydrogenated C$_{60}$, but gradually fade with increasing hydrogenation (Figure~\ref{comp}).  
It is therefore possible that the astronomical C$_{60}$ bands reported in the literature could actually arise in part from slightly hydrogenated C$_{60}$.  { This suggests that  C$_{60}$ and its hydrogenated derivatives may be more prevalent than previously believed as progressive hydrogenation would make the four infrared bands more difficult to detect.}

In Figure~\ref{4peak}, we quantitatively compare the variations of the intrinsic strengths of the four bands as functions of the $m$ values.  It is shown that the total { intrinsic strengths} decrease with increasing hydrogenation, and the { intrinsic strengths} of individual bands have different $m$ dependencies.
The bands at longer wavelengths appear to weaken more rapidly compared to the 7.0\,$\mu$m feature. With increasing hydrogenation, the intensity of the 7.0\,$\mu$m band decreases by a factor of about 1.5 and then remains nearly constant until $m>14$, while the intensities of the 18.9 and 17.4\,$\mu$m bands  continually decrease by more than one order of magnitude. Therefore, the relative strengths of the four skeletal bands are different for C$_{60}$ and 
C$_{60}$H$_m$.

The observed relative strengths  of the four C$_{60}$ bands have been used to infer excitation mechanisms of the molecule \citep{bc12,gar12,rs12, zk13}.
However, it is difficult to distinguish between fluorescence and thermal excitation mechanisms  because of  the uncertainties in the intrinsic band strengths.  This problem was recently re-examined by \citet{bg16}, who presented new laboratory measurements of the intrinsic strengths. Even using  the latest data, the observed band ratios are too dispersed to be assigned to either fluorescence or thermal excitation models.
\citet{bg16} attributed this inconsistency to  contamination of the band strengths from UIE or other emissions, which if true would severely limit the utility of { C$_{60}$ infrared bands as a probe} of physical conditions in circumstellar environments.

In Figure~\ref{exc} we investigate the possibility of whether the hydrogenation of C$_{60}$ may play a role in  the wide scatter of the observed band ratios.  
The observed band ratios are taken from  \citet{bc12}, \citet{gar12}, and \citet{ots14}.  We should note that even for the strong C$_{60}$-source Tc\,1, different authors report different 7.0 $\mu$m/8.5 $\mu$m band ratios partly due to different ways of subtracting other line contributions.  
The variation of the band ratios as a function of temperature under the  thermal excitation model is plotted.  We note that the predicted band ratios by fluorescence model are very close to those by thermal model \citep{bg16}.  
Figure~\ref{exc} shows that the large scatter of the observed 17.4\,$\mu$m/18.9\,$\mu$m band ratios cannot be accounted by the model characterized by single parameter (i.e. average photon energy and temperature for fluorescence and thermal model, respectively).  
However, the inconsistency can be explained by hydrogenation effects. First, different hydrogenation degrees and diverse isomers of C$_{60}$H$_m$ can contribute to the scatter of the observed band ratios.   Secondly, most of C$_{60}$H$_m$ show larger theoretical 17.4\,$\mu$m/18.9\,$\mu$m band ratios than C$_{60}$.
The fact that the observed data lie within the range covered by the theoretical values of C$_{60}$H$_m$ in Figure~\ref{exc} suggests that  hydrogenated C$_{60}$ may be present in these sources and contribute to the observed infrared bands.

\subsection{Search for circumstellar C$_{60}$H$_m$}

One of the major challenges in searching for circumstellar C$_{60}$H$_m$ is the large number of isomers. Although C$_{60}$H$_{36}$ and C$_{60}$H$_{18}$  are the dominant products of the experiments of C$_{60}$ exposed to hydrogen atoms, the physical conditions of circumstellar envelopes are very different from those in a terrestrial laboratory.  Given the very low densities in circumstellar envelopes, the less stable fulleranes are likely to be  present. Through averaging the theoretical spectra of  different isomers (Figure~\ref{comp}), the bands that are commonly present in all the isomers can serve as guides for the search of circumstellar C$_{60}$H$_m$.
One of the distinctive features of the fullerane spectra is the emergence of a 13--17\,$\mu$m plateau (Figure~\ref{comp}).  As the  number of H atoms increases, a band at 15\,$\mu$m  becomes more prominent.  
The variation of strengths of the 15\,$\mu$m  and the 13--17\,$\mu$m plateau features with the $m$ values are illustrated in Figures~\ref{15m} and \ref{15m_pla}. 
We can see that the plateau feature is generally stronger  than the 15\,$\mu$m band except for C$_{60}$H$_{36}$. For heavily hydrogenated C$_{60}$, the 15\,$\mu$m band clearly stands out from the plateau.

The 15 $\mu$m feature is unique to fulleranes.  There is no known aromatic molecules having strong bands at 15\,$\mu$m. 
The bending modes of aliphatic compounds are concentrated around 8 and 12 $\mu$m.
CO$_2$ has a band at this wavelength, but it is not expected in carbon-rich environments.  Thus the 15\,$\mu$m band might serve as a useful probe of heavily hydrogenated C$_{60}$ in circumstellar envelopes.

We have undertaken a search for the 15\,$\mu$m band in known C$_{60}$ sources (Appendix A).  
The exact positions of the vibrational bands in theoretical spectra are affected by the scaling factor applied (section \ref{method}) and this is evident in the slight wavelength differences seen in the theoretical and experimental wavelengths spectra of C$_{60}$ (Figure~\ref{com_c60}). 
Consequently, we cannot take the peak wavelength of  15\,$\mu$m as the definitive signature of fullerane . 
Nevertheless, we note that several objects display features with wavelengths close to 15\,$\mu$m.
Three C$_{60}$-containing objects (IRAS 01005+7910, IRAS 06338+5333, NGC 7023) display a feature at 15\,$\mu$m. 
\footnote{{ We note that the 8.5\,$\mu$m features in the three sources are badly blended with at 8.6\,$\mu$m UIE band and the 7 $\mu$m features are also blended with H$_{2}$ S(5) at 6.91 $\mu$m line and thus the three sources are not included in Fig.~\ref{exc}.}
}
We note that none of these three objects have a hot central star, which is consistent with the suggestion of \citet{dg16} that fulleranes are quickly destroyed by increasing UV radiation. Given the weakness of the features, we must regard the detections in the three sources as tentative.

Since hydrogenation can greatly suppress the four C$_{60}$ skeletal bands, the search for fullerane should not be limited to sources showing the four C$_{60}$ bands. A more extensive search for the 15 $\mu$m band should be conducted for a wider range of objects.

In the case of NGC\,7023, spatially resolved spectra are available (Appendix B).  
Figure~\ref{ngc7023} shows 5 spectra of NGC 7023, from closest to the central star (`S1', top panel) to farthest from the central star (`S5', bottom panel).   We can see that the 15\,$\mu$m feature appears in positions `S3' and `S4', where the C$_{60}$ emission is becoming weak and H$_2$ emission is appearing.
 \citet{sel10} have shown that, compared to H$_2$ emission, the 18.9\,$\mu$m C$_{60}$ feature lies closer to the central star where
UV radiation is more intense, while the UIE bands peak between the C$_{60}$ and H$_2$ emission. 
Figure~\ref{ngc7023} indicates that the 15\,$\mu$m feature and UIE are roughly co-spatial, supporting the idea that the preferential conditions for the presence of fulleranes are  H atom rich and void of intense UV radiation.
{ The observed feature is relative narrow, suggesting that C$_{60}$H$_m$ isomers, if exist, are not very diverse, and the physical conditions favor the formation of specific chemical structures.}

We note that \citet{ber13} detected C$_{60}^+$ in the spectrum at 7.5$\arcsec$ north-west from the central star of NGC 7023 (position 2 of their Figure 1).
Figure \ref{map} shows that the C$_{60}^+$, C$_{60}$, UIE, and H$_2$ emission regions are spatially segregated in NGC 7023, suggesting that photochemsitry plays a role in the formation of fullerenes. 
In the hypothesis of PAH being the carrier of UIE bands, fullerenes are formed in a top-down process where   large PAHs are dehydrogenated by UV radiation, followed by carbon atom ejection and isomerisation \citep{ber12}. 
{ If our detection and identification of the 15\,$\mu$m feature is correct,} this  scenario is contradicted by our observations of the 15\,$\mu$m fullerane feature being coincident with the distribution of the UIE bands.
Alternatively,  fullerenes/fulleranes are formed by the photochemical effects on  mixed amorphous aromatic/aliphatic compounds  \citep{gar10,mic12,zk13}.  More theoretical work is needed to further pursue this idea.

\section{Summary}

Fulleranes are potentially an important constituents of the interstellar medium for their possible role in the formation of molecular hydrogen and their relationships to the carriers of the UIE and DIB phenomena \citep{ca10}.  There are also strong theoretical and experimental reasons to believe that fulleranes are widely present in the circumstellar and interstellar environments.   
However, the search for circumstellar  fulleranes has so far been hindered by the lack of detailed knowledge of their spectra. In this paper
we performed a  theoretical investigation of  the vibrational spectra of fulleranes using quantum chemistry methods.
The results show that hydrogenation can have a significant effect on the determinations of abundance and excitation mechanism of C$_{60}$. 
The  inconsistent C$_{60}$ band ratios between observations and model predictions might be  an indirect proof for the existence of
slightly hydrogenated C$_{60}$. 

A feature at 3.4 $\mu$m due to C--H stretching modes is generally present in the spectra of all fulleranes.  
Also noticeable are broad features around 8, 15, and 20 $\mu$m, which arise from C--H bending modes, carbon skeleton radial modes, and carbon skeleton deformation modes, respectively.  The strength of the 15\,$\mu$m increases with increasing hydrogenation might possibly be a tracer  of fulleranes.
A spectral search suggests that this feature is present in three C$_{60}$-containing nebulae with no strong UV radiation background.

One of the major challenges in the study of astronomical fulleranes is the large (up to $10^{15}$ as shown in Figure \ref{isonum}) number of isomers, making theoretical calculations on all isomers impossible.  In this paper, we have limited our study to isomers of lowest energies.  Our results show definite spectral trends with progressive hydrogenation.  
The astronomical implications on the C$_{60}$ skeletal bands are therefore likely to be independent  on the selection of specific isomers.
We plan to extend our computation to more fulleranes including different numbers of carbon and hydrogen atoms as well as
different bond arrangements. 

\acknowledgments

We gratefully acknowledge Franco Cataldo for providing us with their experimental spectra of fullerene. We also thank an anonymous referee and
Anibal  Garc{\'{\i}}a-Hern{\'a}ndez for helpful comments.
This work  makes use of  observations made with the {\it Spitzer Space Telescope}, which is operated by the Jet Propulsion Laboratory, California Institute of Technology, under a contract with NASA.  
Financial support for this work was provided by the Research Grants Council of the Hong Kong under grants  HKU 7027/11P and HKU7062/13P.

\newpage

\appendix

\section{A search for the 15\,$\mu$m feature in C$_{60}$ sources}

We examined the infrared spectra of 35 sources that have been reported to contain C$_{60}$ emission to search for the fullerane 15\,$\mu$m feature.  The C$_{60}$ source list (Table 1) includes  young stellar objects (YSOs), Herbig Ae/Be star (HAe/Be), planetary nebulae (PNs), post-asymptotic giant branch stars (PAGBs), proto planetary nebula (PPN), R Coronae Borealis star (RCB), binary, and reflection nebula (RN).
The objects are classified as carbon-rich (C) or mixed chemistry (M) (column 4 of Table 1) based on other spectral features in the infrared spectra.
The spectra were extracted from the {\it Infrared Space Observatory (ISO)} and {\it Spitzer} archives.  

Three objects (IC 418, HR 4049, and NGC 7023) in our sample were observed between 1996 June and 1998 February using {\it ISO} {\it Short-Wavelength Spectrometer (SWS)}. The observations were performed using the Astronomical Observation Template (AOT) 01 mode at various speeds with spectral resolving power ($\lambda$/$\triangle\lambda$) ranging from 500 to 1600, covering a wavelength range from 2.4--45.2 $\micron$. The aperture sizes are 14$\arcsec\times$ 20$\arcsec$ for 2.4--12 $\micron$ region and 14$\arcsec\times$ 27$\arcsec$ 
for 11--28 $\micron$ band in SWS module, respectively, and the exposures on sources of the measurements varied between 1834\,s and 6538\,s, depending on the source brightnesses. The observations for all objects are centered on the core positions.  The data reduction and calibration were performed through the  standard procedure.

The mid-infrared spectra of the other 32 objects in Table 1 are obtained with the {\it Infrared Spectrograph} (IRS) on board the {\it Spitzer Space Telescope}. All of these objects were observed using the Short-Low (SL) and Short-High (SH) modules, covering the wavelength coverage 5.2--19.6 $\micron$ with spectral dispersion of $R\sim$ 64--600. The aperture sizes are 3$\farcs$6 $\times$ 57$\arcsec$ and 4$\farcs$7 $\times$ 11$\farcs$3 in SL and SH modules, respectively, and the total integration times of IRS observations range from 39\,s to 2462\,s, depending on the sources' expected mid-infrared fluxes. Data were reduced starting with basic calibrated data from the Spitzer Science Center's pipeline version s18.7 and were run through the {\bf IRSCLEAN} program to remove rogue pixels. Then we employed point-source spectral extractions to extract spectra using the Spectral Modeling, Analysis and Reduction Tool. A final spectrum was produced using the combined IRS observations to improve the S/N.

Figure~\ref{spectra} shows the spectra of the 35 objects.  The continua were fitted with a third-order polynomial and  subtracted from the spectra.  The positions of the four C$_{60}$ bands and the 15\,$\mu$m feature are marked with vertical lines. The search results are given in column 3 of  Table~\ref{tab1}, where ticks indicate tentative detections and colons indicate marginal or uncertain detections.

\section{The spatially resolved spectra of NGC\,7023}

Infrared spectroscopic observations of NGC\,7023 were carried out by the IRS  onboard the {\it Spitzer Space Telescope} in the spectral mapping mode on 2004 August (AOR 3871232). We extracted five high-resolution spectra at different positions, which were taken with the SH module covering the wavelength range 9.9--19.6\,$\mu$m.  
The slit sizes are $4\farcs7\times11\farcs3$ and the exposure time of each spectrum is 20\,s. 
Data were reduced using basic calibrated data from the Spitzer Science Center's pipeline version s18.8 and the IRSCLEAN program for removing rogue pixels was performed. A low-order polynomial fit was applied to subtract the continuum.  These spatially resolved spectra are shown in Figure~\ref{ngc7023}.

We note that the 15 $\mu$m feature is detected in both the {\it ISO} (Figure \ref{spectra}) and {\it Spitzer} (Figure \ref{ngc7023}) spectra of NGC 7023.

\section{Online database}
A database of infrared spectra of fulleranes is presented as online supplementary material.

\begin{deluxetable}{lcc@{\extracolsep{0.1in}}ccc}
\tabletypesize{\scriptsize} 
\tablecaption{The 15\,$\mu$m feature in C$_{60}$ sources} \tablewidth{0pt}
\tablehead{\colhead{Name} & \colhead{Instrument} & \colhead{15.0\,$\micron$} & \colhead{Class$^{a}$} & \colhead{Object Type} & 
\colhead{Reference} 
}
\startdata
2MASS J06314796+0419381 & Spitzer IRS & ... & M & YSO & (1) \\
DY Cen & Spitzer IRS & ... & C & RCB & (2) \\
HD 52961 & Spitzer IRS & $:^{b}$ & M & PAGB & (3) \\
HD 97300 & Spitzer IRS & ... & C & HAe/Be & (1) \\
Hen 2-68 & Spitzer IRS & $:^{c}$ & C & PN & (4) \\
HR 4049 & ISO SWS & $:^{b}$ & M & PAGB & (1) \\
IC 418 & ISO SWS & $:^{c}$ & C & PN & (4) \\
IRAS 01005+7910 & Spitzer IRS & $\surd$$^{d}$ & C & PPN & (5) \\
IRAS 06338+5333 & Spitzer IRS & $\surd$$^{d}$ & M & PAGB & (3) \\
ISOGAL-P J174639.6-284126 & Spitzer IRS & $:^{e}$ & M & YSO & (1) \\
K 3-54 & Spitzer IRS & ... & C & PN & (6) \\
K 3-62 & Spitzer IRS & ... & C & PN & (4) \\
Lin 49 & Spitzer IRS & ... & C & PN & (7) \\
LMC 02 & Spitzer IRS & ... & C & PN & (6) \\
LMC 25 & Spitzer IRS & ... & C & PN & (6) \\
LMC 48 & Spitzer IRS & ... & C & PN & (6) \\
LMC 99 & Spitzer IRS & ... & C & PN & (6)\\
M 1-6 & Spitzer IRS & ... & C & PN & (4) \\
M 1-9 & Spitzer IRS & ... & C & PN & (4) \\ 
M 1-11 & Spitzer IRS & $:^{d}$ & C & PN & (4) \\ 
M 1-12 & Spitzer IRS & ... & C & PN & (4) \\ 
M 1-20 & Spitzer IRS & ... & C & PN & (4) \\ 
M 1-60 & Spitzer IRS & ... & C & PN & (6) \\ 
NGC 7023 & ISO SWS & $\surd$$^{d}$ & C & RN & (8) \\ 
         & Spitzer IRS & $\surd$$^{d}$ & C & RN & (8)  \\
SaSt 2-3 & Spitzer IRS & ... & C & PN & (4) \\ 
SMC 13 & Spitzer IRS & ... & C & PN & (6) \\ 
SMC 15 & Spitzer IRS & ... & C & PN & (6) \\
SMC 16 & Spitzer IRS & ... & C & PN & (6) \\
SMC 18 & Spitzer IRS & ... & C & PN & (6) \\
SMC 20 & Spitzer IRS & ... & C & PN & (6) \\ 
SMC 24 & Spitzer IRS & ... & C & PN & (6) \\ 
SMC 27 & Spitzer IRS & ... & C & PN & (6) \\ 
SSTGC 372630 & Spitzer IRS & ... & M & YSO & (1) \\ 
Tc 1 & Spitzer IRS   & $:^{c}$ & C & PN & (9) \\ 
XX Oph & Spitzer IRS & ... & M & Binary & (10) \\

\enddata
\label{tab1}
\tablenotetext{{\it a}}{M: mixed-chemistry dust, C: carbon-rich dust.}
\tablenotetext{{\it b}}{A strong peak at 14.95\,$\micron$ and two weak ones at about 15.05 and 15.1\,$\micron$.}
\tablenotetext{{\it c}}{An extreme weak peak around 15.1\,$\micron$.}
\tablenotetext{{\it d}}{A weak but still visible peak at 15.0\,$\micron$.}
\tablenotetext{{\it e}}{A visible peak at 14.9\,$\micron$.}

\tablerefs{(1) Roberts et al. (2012); (2) Garc\'{i}a-Hern\'{a}ndez et al. (2011); (3) Gielen et al. (2011); (4) Otsuka et al. (2014); 
(5) Zhang \& Kwok (2011); (6) Garc\'{i}a-Hern\'{a}ndez et al. (2012); (7) Otsuka et al. (2016); (8) Sellgren et al. (2010);  
(9) Cami et al. (2010); (10) Evans et al. (2012).}

\end{deluxetable}

\begin{figure*}
\epsfig{file=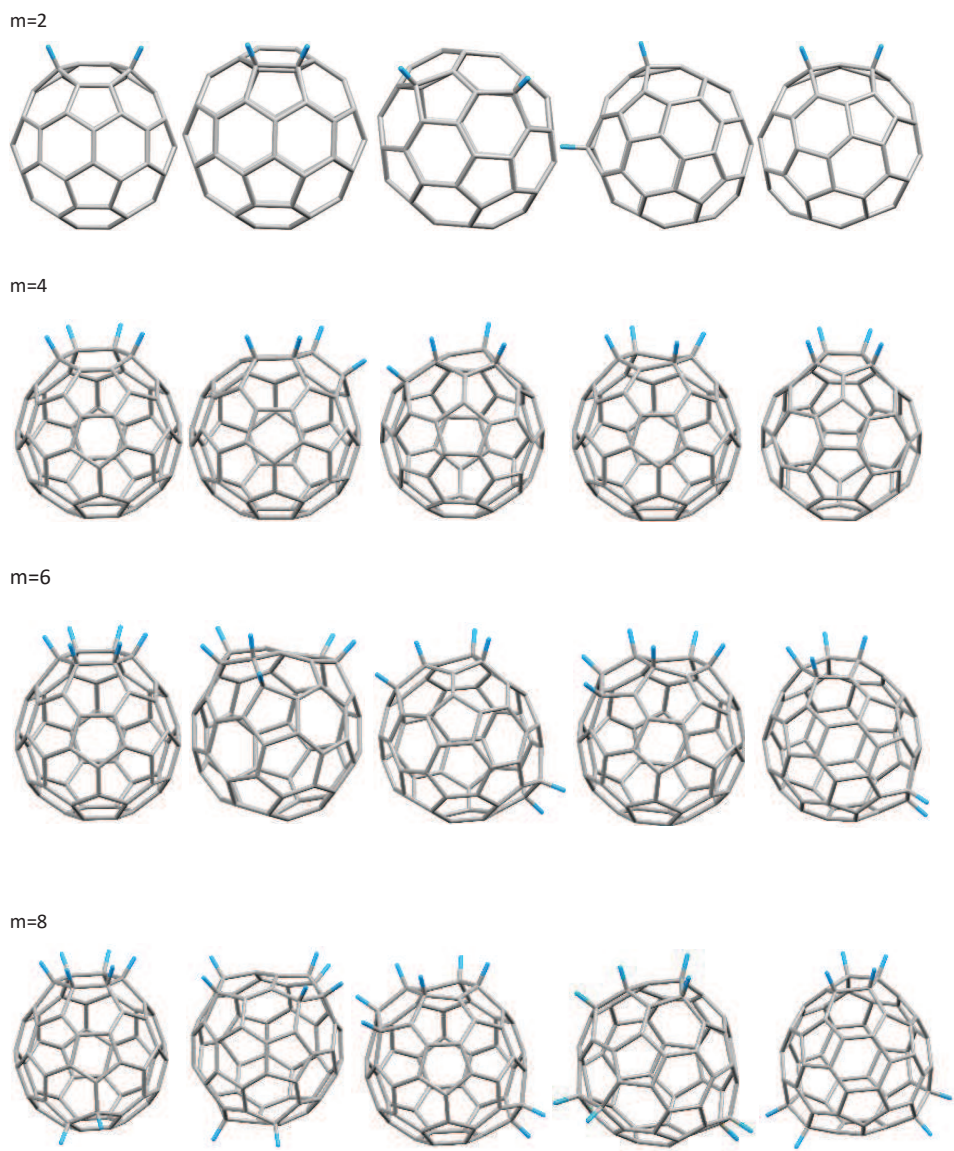, height=20cm}
\caption{
Local minimum geometries of C$_{60}$H$_m$ isomers from model calculations.  
A total of 55 structures, 5 isomers for each of C$_{60}$H$_m$, $m=2, 4, 6, 8, 10, 12, 14, 16, 18, 20, 36$ are included in our study.  The numbers of the isomers are assigned as 1 to 5 from left to right.  The carbon cage is shown in grey and the C$-$H bonds are shown in blue.  We can clearly see the geometric distortion of the C$_{60}$ cage when hydrogen atoms are attached.}
\label{structure}
\end{figure*}

\begin{figure*}
\setcounter{figure}{1}
\epsfig{file=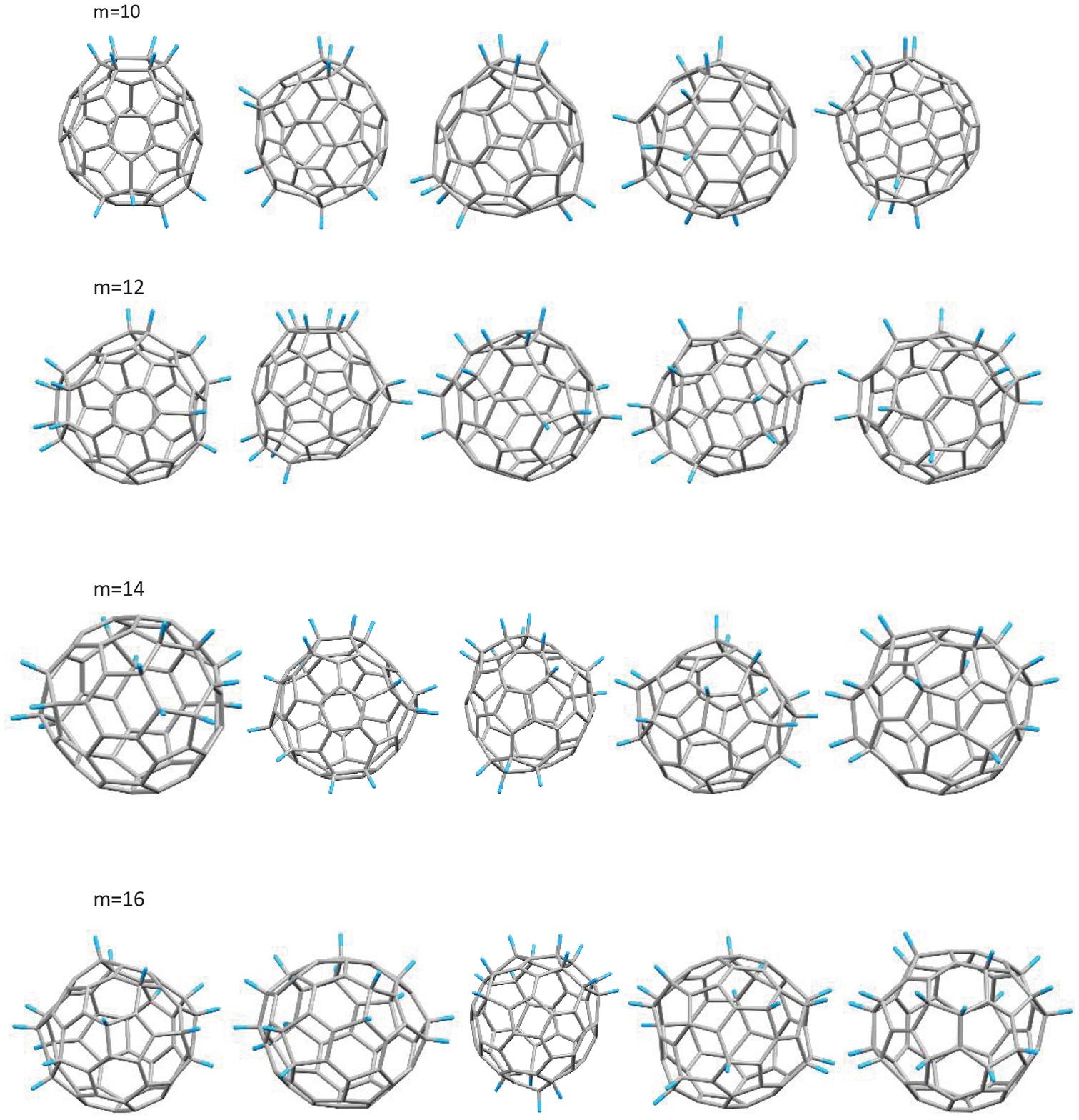, height=22cm}
\caption{continued.}
\end{figure*}

\begin{figure*}
\setcounter{figure}{1}
\epsfig{file=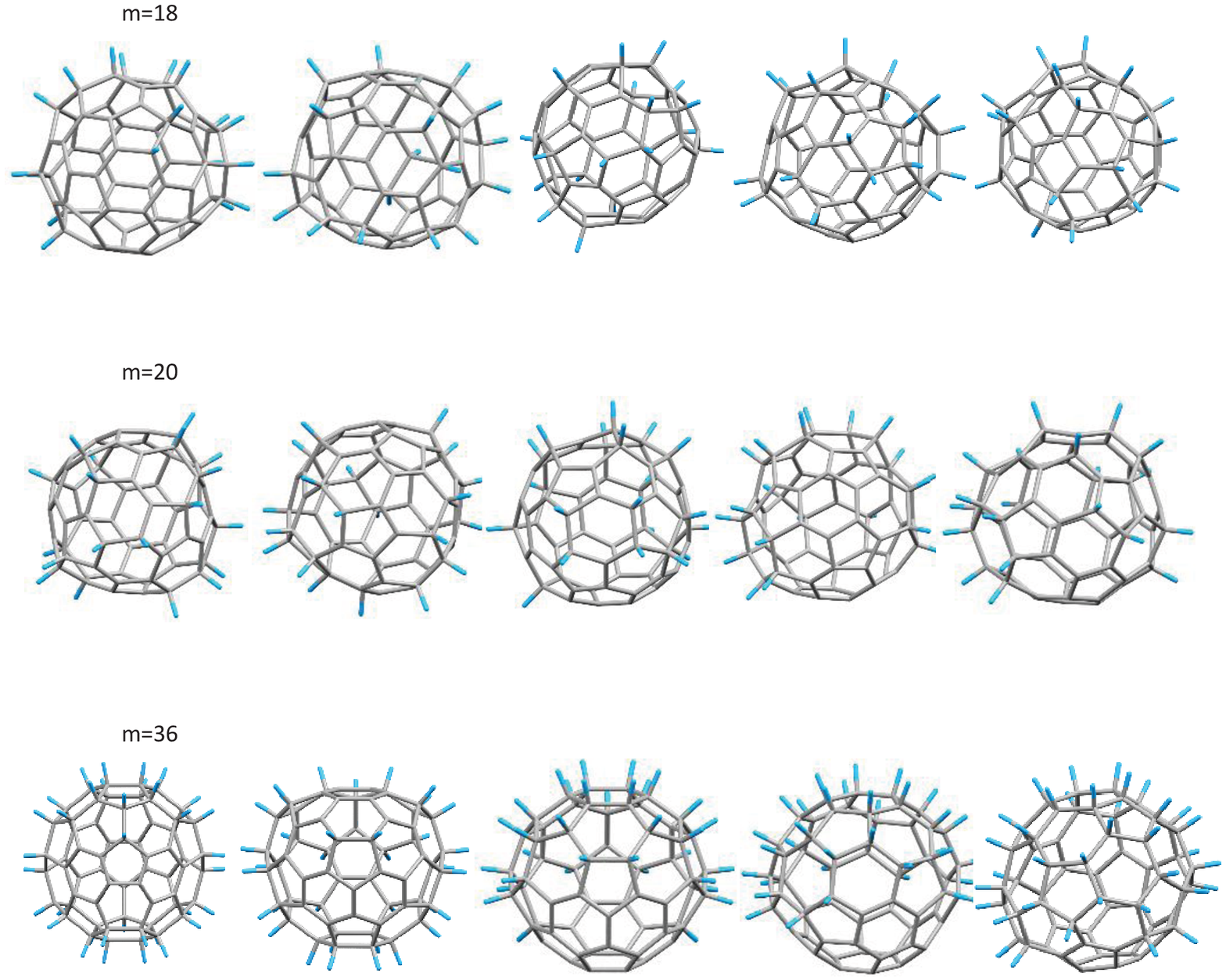, height=22cm}
\caption{continued.}
\end{figure*}

\begin{figure*}
\epsfig{file=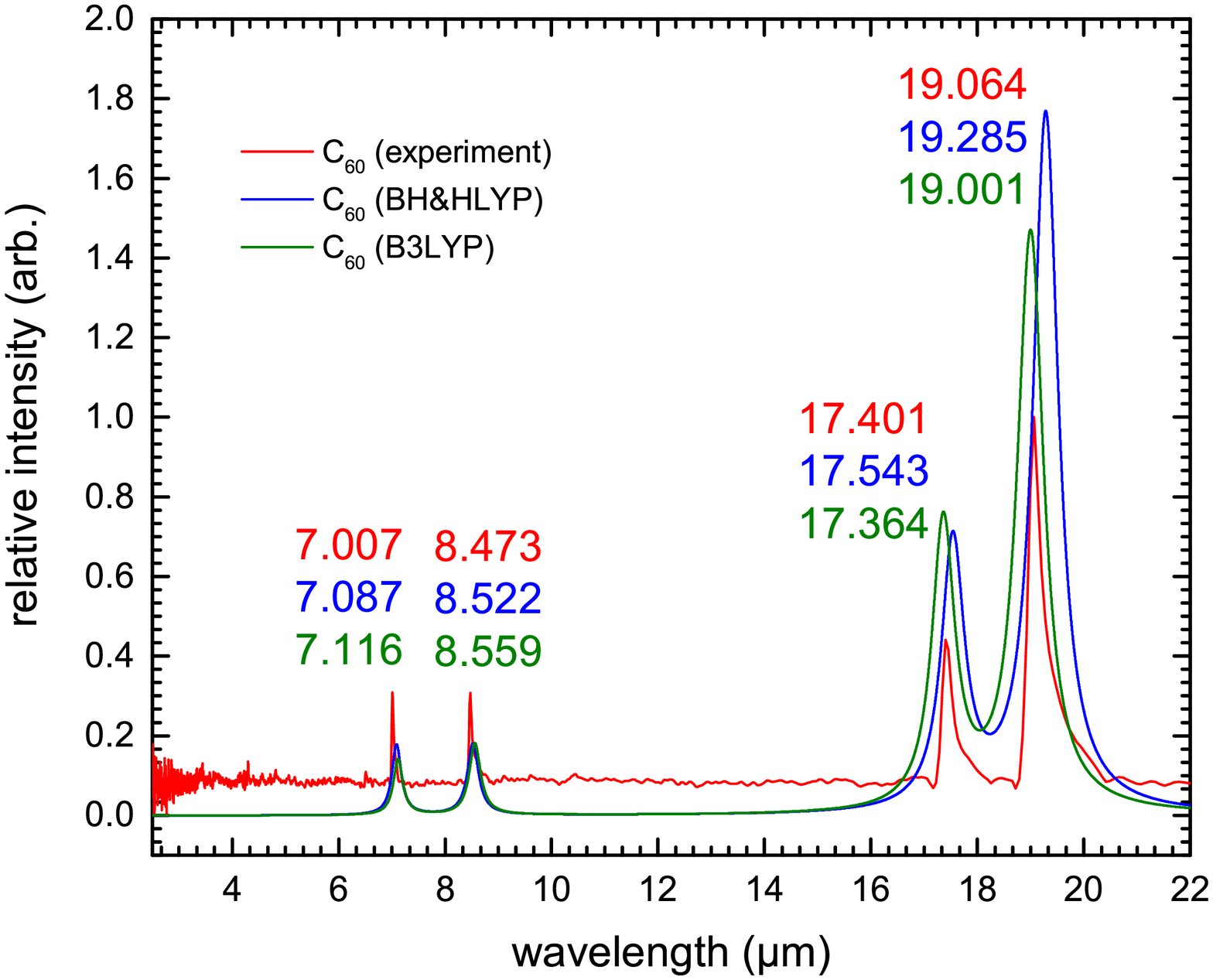, height=13cm}
\caption{
Theoretical spectra of C$_{60}$ obtained using the B3LYP (green) and BH\&HLYP (blue) functionals, compared with the experimental data (red) of Cataldo (private communication). The peak wavelengths are indicated with the corresponding colors.
}
\label{com_c60}
\end{figure*}

\begin{figure*}
\epsfig{file=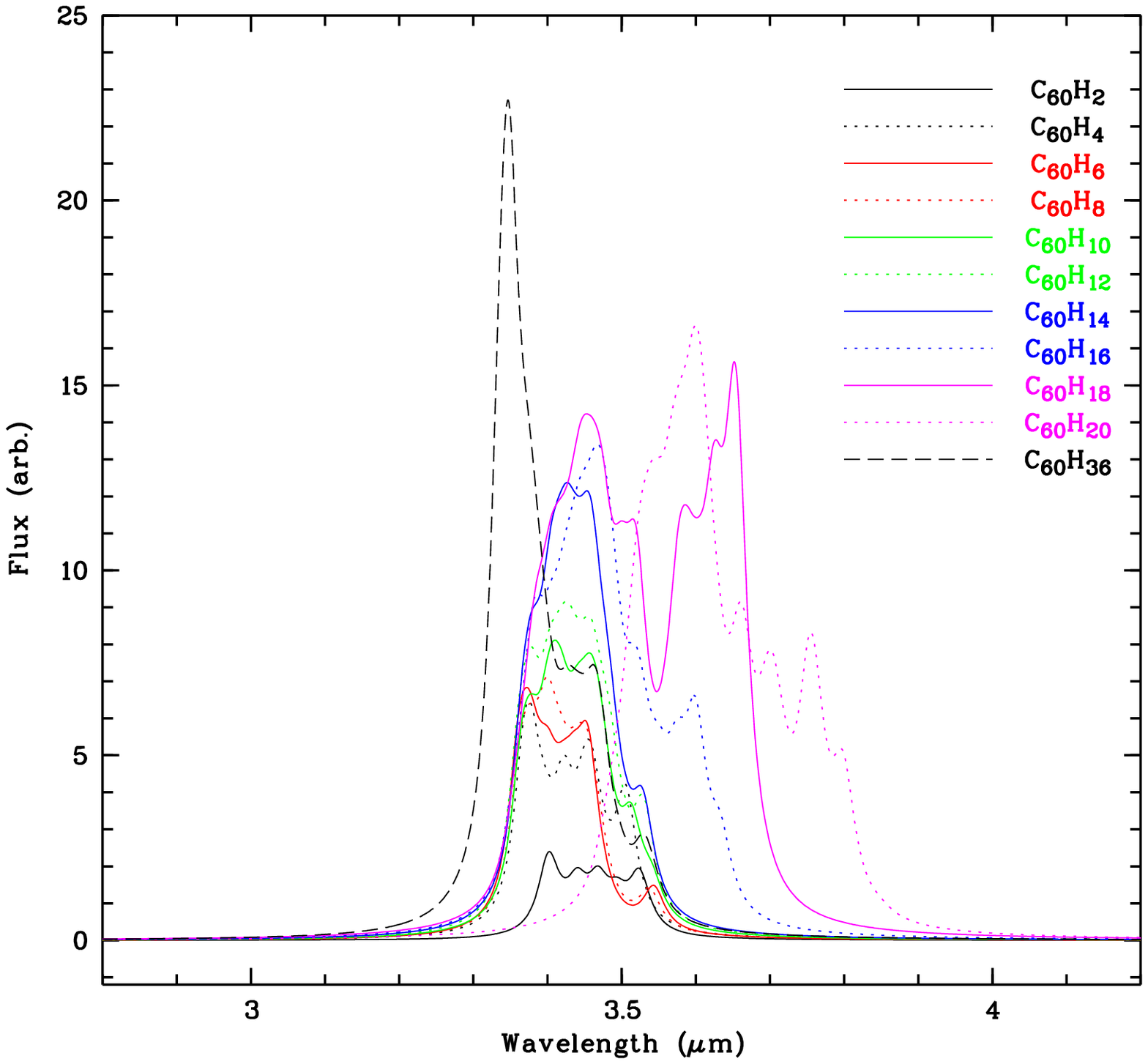, height=13cm}
\caption{Theoretical profiles of the C$_{60}$H$_m$ 3.4\,$\mu$m feature. Each profile is the mean of five isomers.}
\label{3um}
\end{figure*}

\begin{figure*}
\epsfig{file=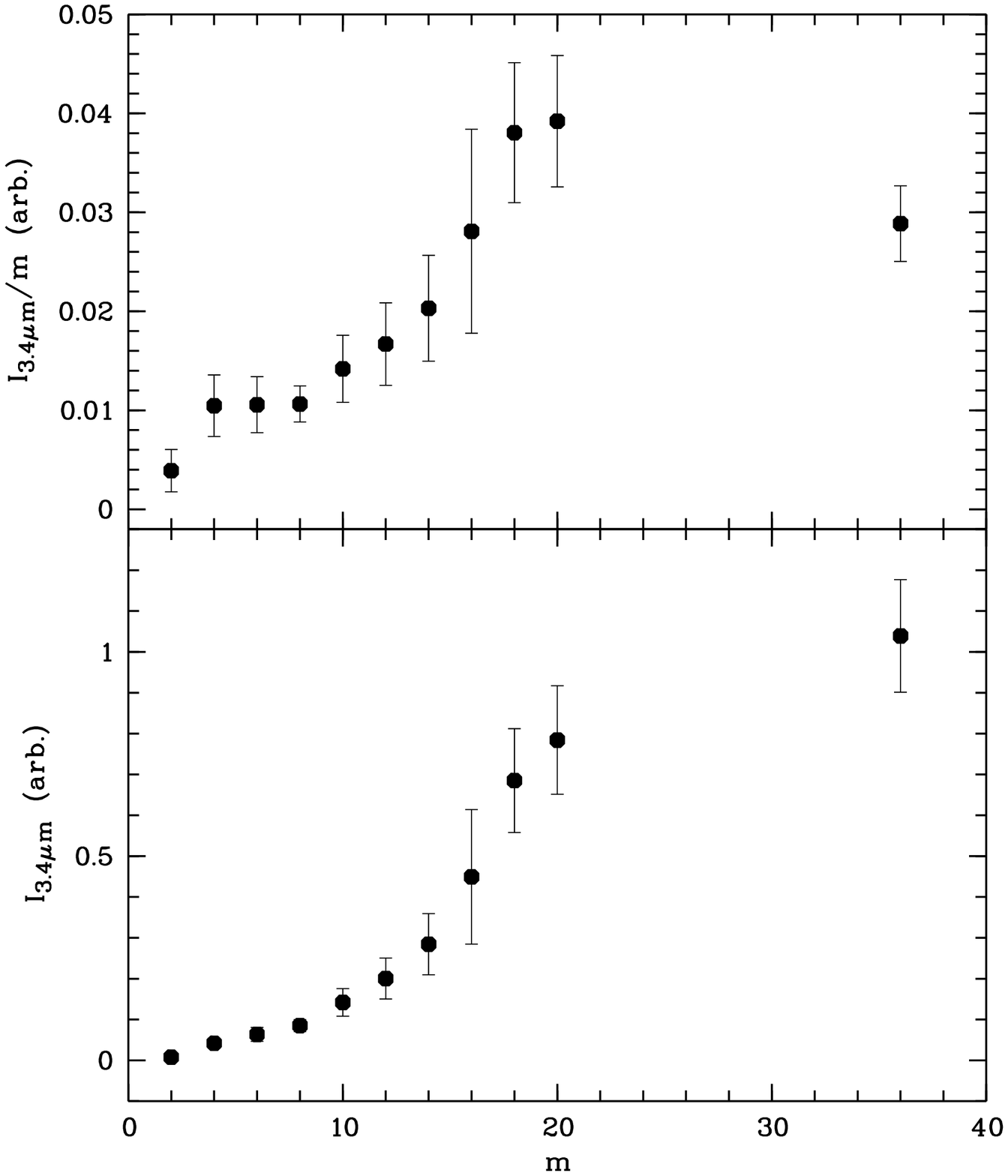, height=15cm}
\caption{ The intensities of the 3.4\,$\mu$m feature  (I$_{3.4\mu{\rm m}}$; lower panel) and I$_{3.4\mu{\rm m}}/m$ (upper panel) versus the $m$ values of  C$_{60}$H$_m$. The bars represent the standard deviations of the means of different isomers.
}
\label{3um_m}
\end{figure*}

\begin{figure*}
\epsfig{file=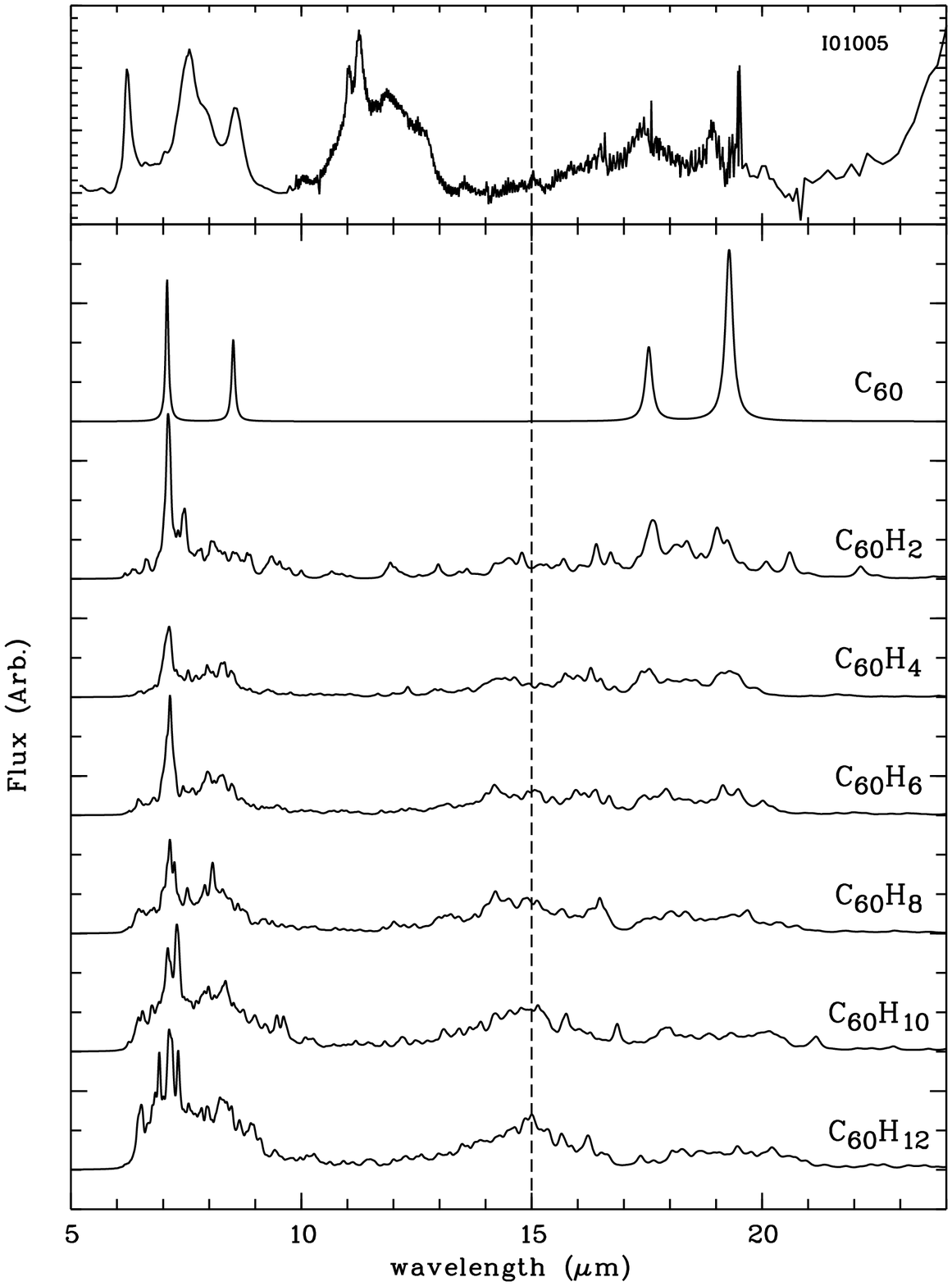, height=20cm}
\caption{The spectrum of IRAS\,01005+7910 (top panel) and theoretical spectra of C$_{60}$ (second panel) and C$_{60}$H$_m$ (lower panels). The dashed line marks the position of the 15\,$\mu$m feature.}
\label{comp}
\end{figure*}

\addtocounter{figure}{-1}
\begin{figure*}
\epsfig{file=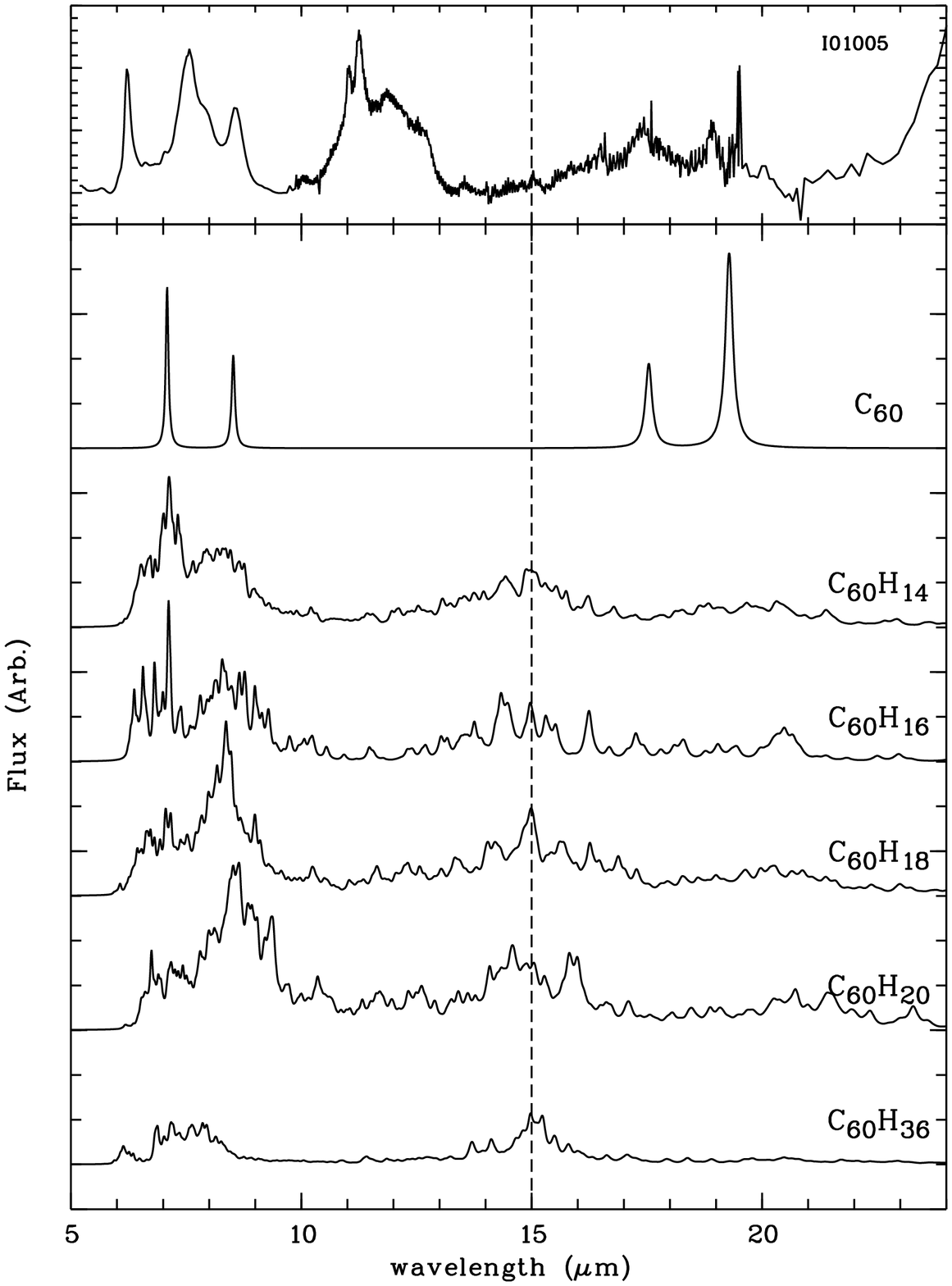,
height=20cm, }
\caption{continued.}
\end{figure*}

\begin{figure*}
\epsfig{file=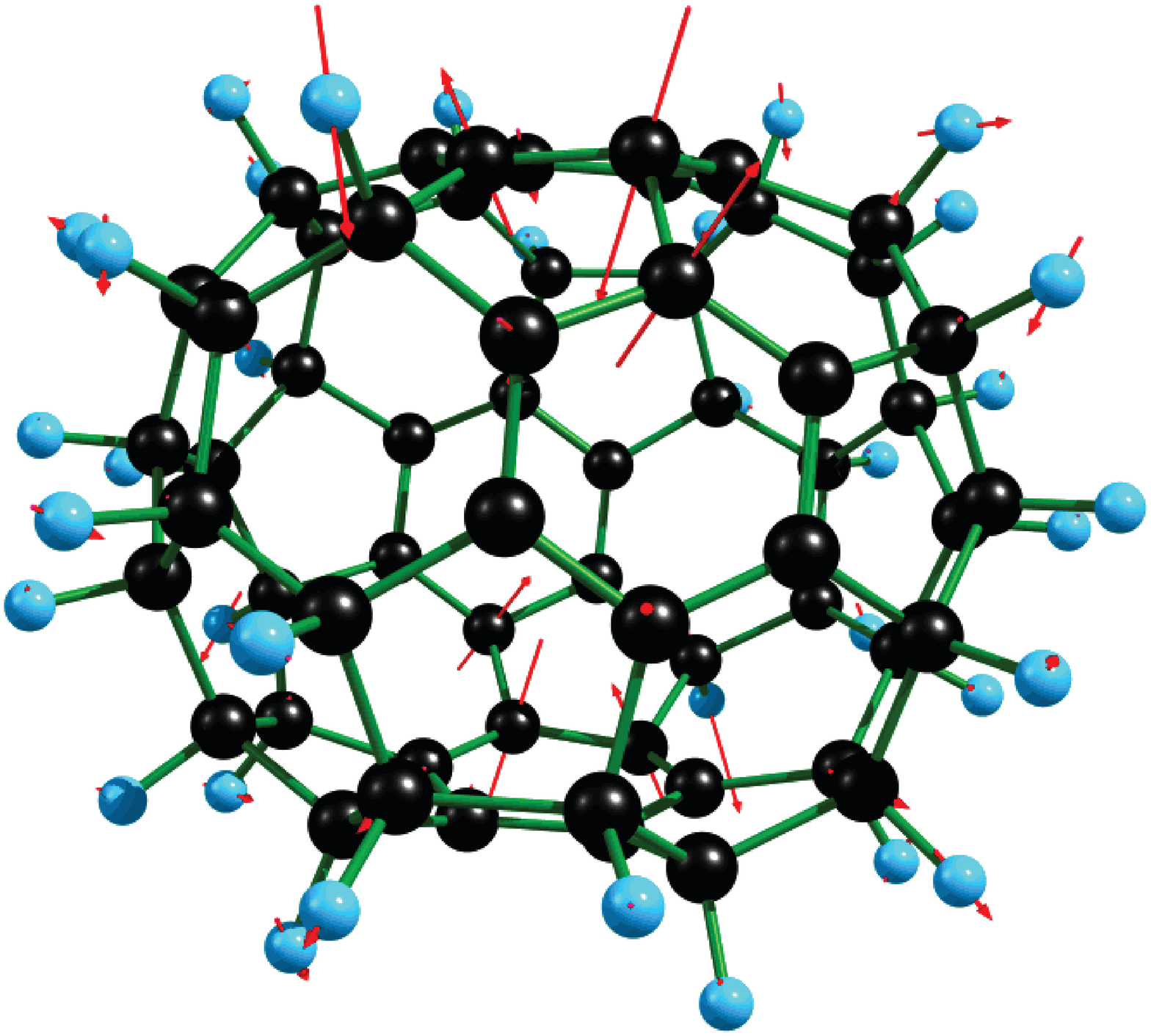, height=10cm}
\caption{The vibrational mode of C$_{60}$H$_{36}$ responsible for the 15\,$\mu$m band. Black and blue circles represent carbon and hydrogen atoms, respectively.
The C$-$C and C$-$H bonds are shown as green lines. The red arrows indicate  the direction of relative motions.  An animation of this vibration mode can be viewed in the on-line version of the paper.
}
\label{stru}
\end{figure*}

\clearpage

\begin{figure*}
\plotone{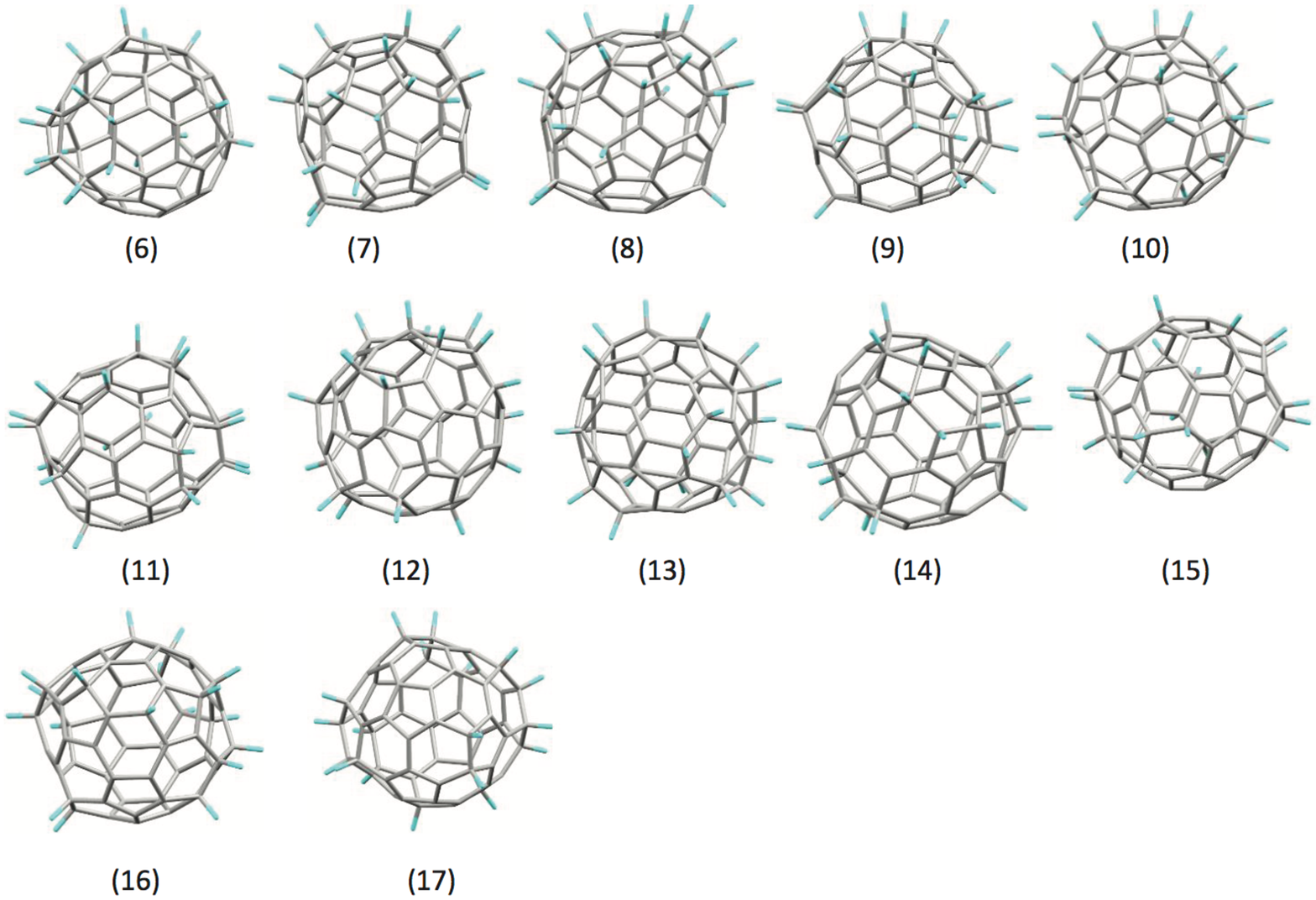}
\caption{Geometry of 15 additional isomers of C$_{60}$H$_{18}$.   The numbers in bracket indicate the isomer numbers.  
}
\label{c60h18geo}
\end{figure*}

\begin{figure*}
\plotone{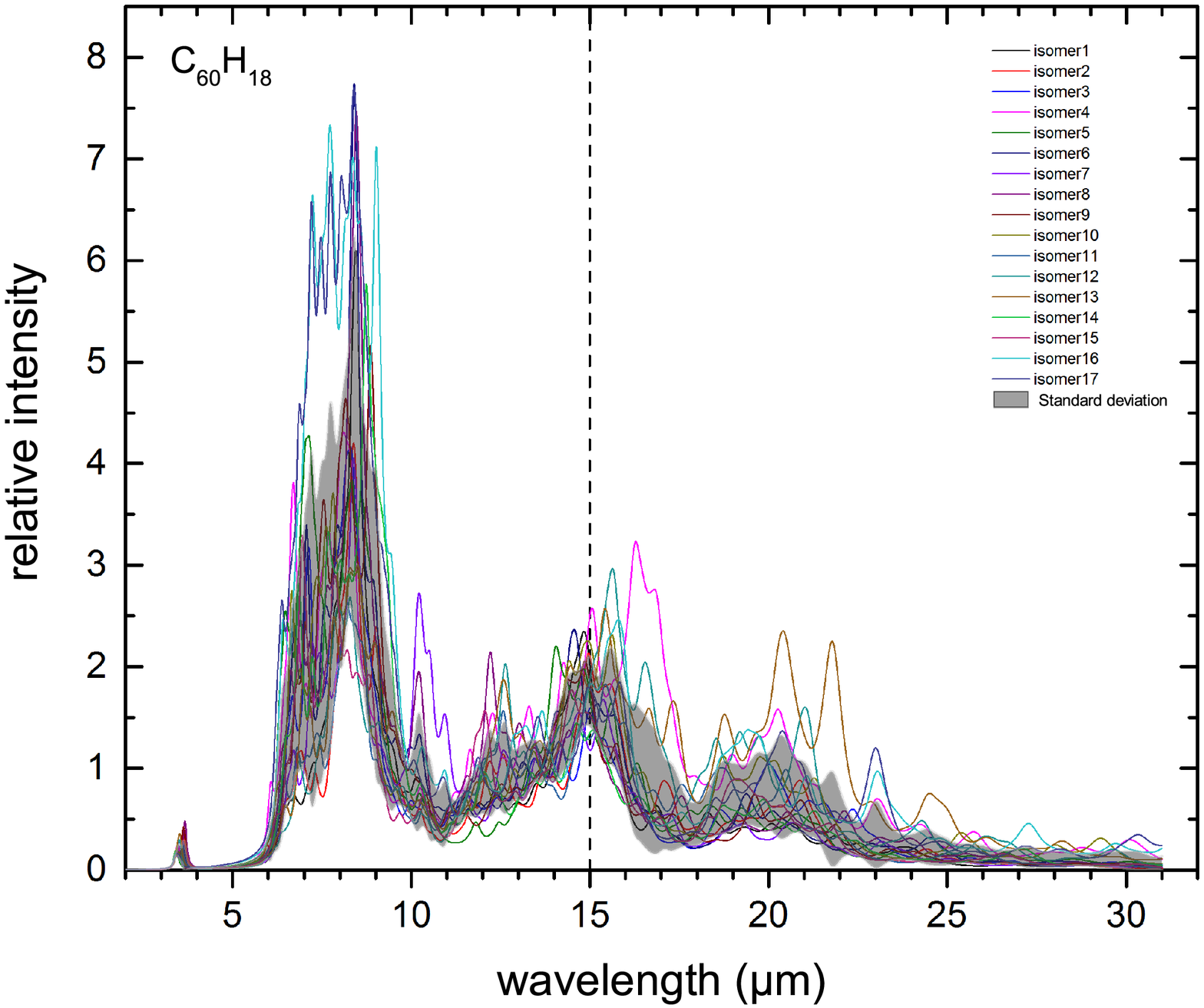}
\caption{Spectral variation of 20 isomers of C$_{60}$H$_{18}$.  The isomer numbers in the legend are those given in Figures \ref{structure} and  \ref{c60h18geo}.  The vertical dashed line marks the position of the 15 $\mu$m feature.  The grey area represents the area covered by one standard deviation from the mean from the 20 spectra.
}
\label{c60h18}
\end{figure*}


\begin{figure*}
\begin{center}
\epsfig{file=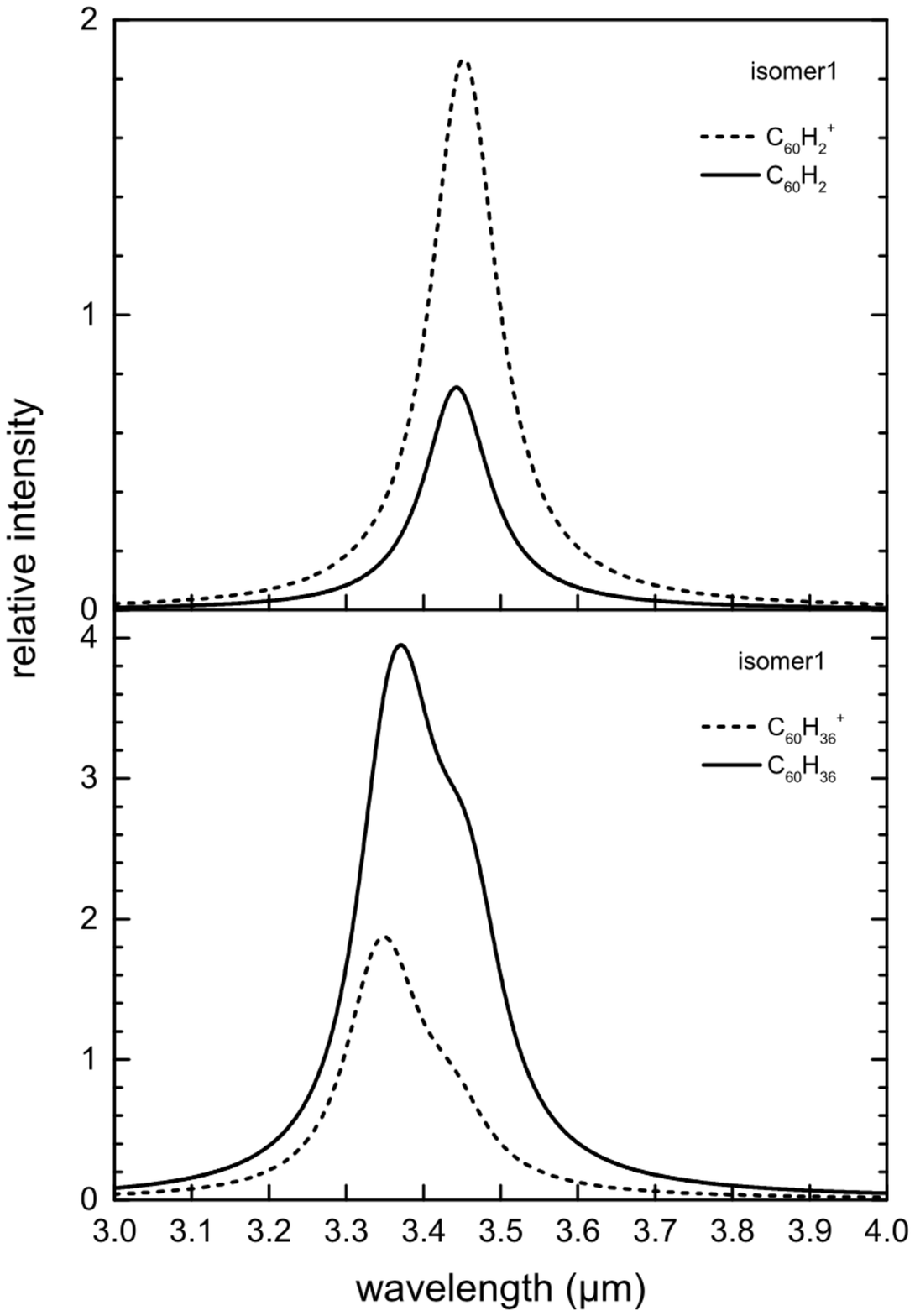, width=10cm}
\caption{Theoretical spectra of two fullerane cations in the 3--4 $\mu$m region.  The upper panel shows the comparison between the spectra of C$_{60}$H$_2$ (solid line) and C$_{60}$H$_2^+$ (dashed line) and the lower panel between C$_{60}$H$_{36}$ (solid line) and C$_{60}$H$_{36}^+$ (dashed line).
}
\label{cations}
\end{center}
\end{figure*}

\clearpage

\begin{figure*}
\epsfig{file=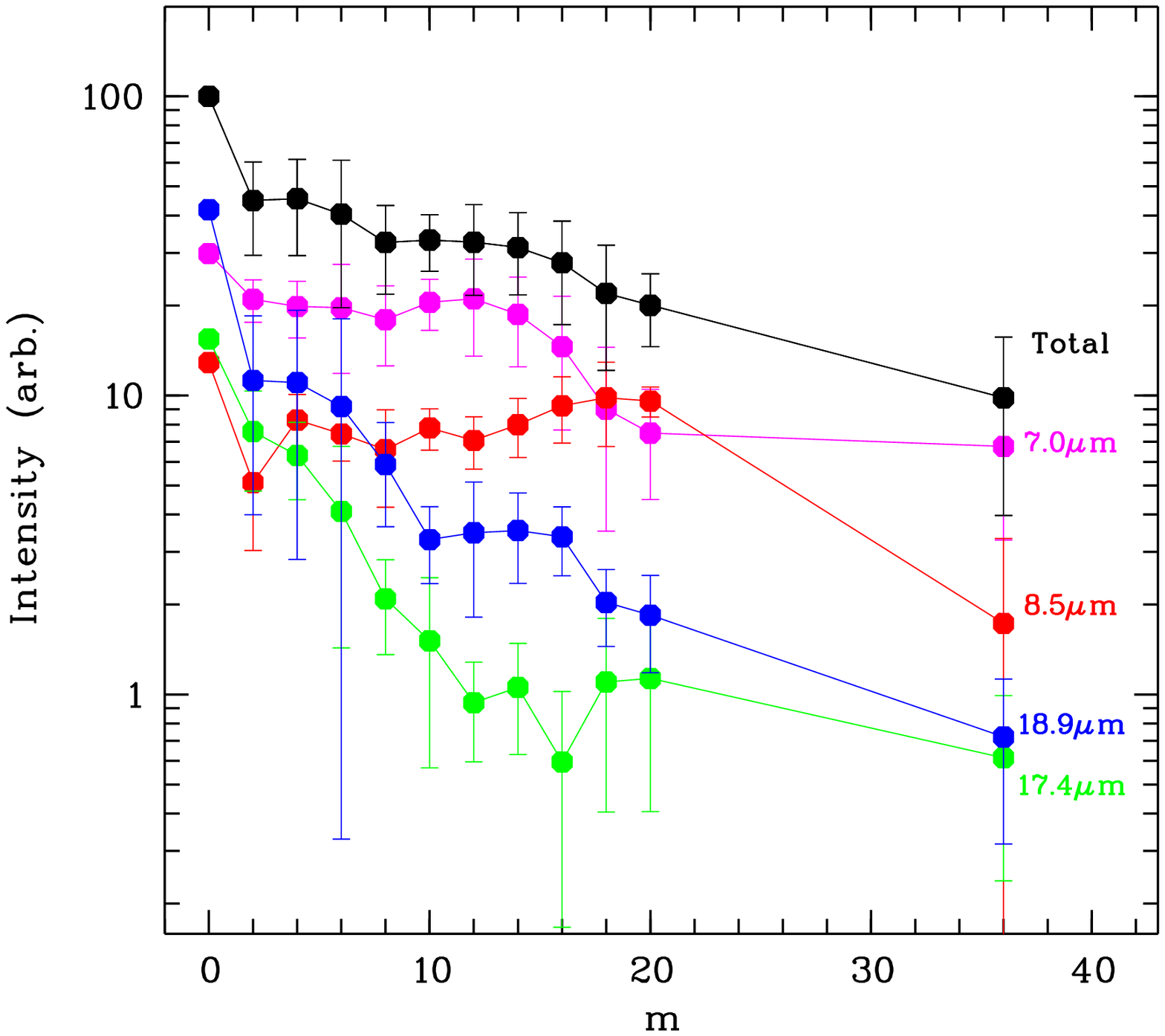, height=13cm}
\caption{The intrinsic strengths of the four carbon skeletal vibration modes (7.0, 8.5, 17.4, and 18.9 $\mu$m) versus the $m$ values of  C$_{60}$H$_m$. The black curve represents the total of the four bands.  The error bars represent the standard deviations of the means of different isomers.
}
\label{4peak}
\end{figure*}

\begin{figure*}
\epsfig{file=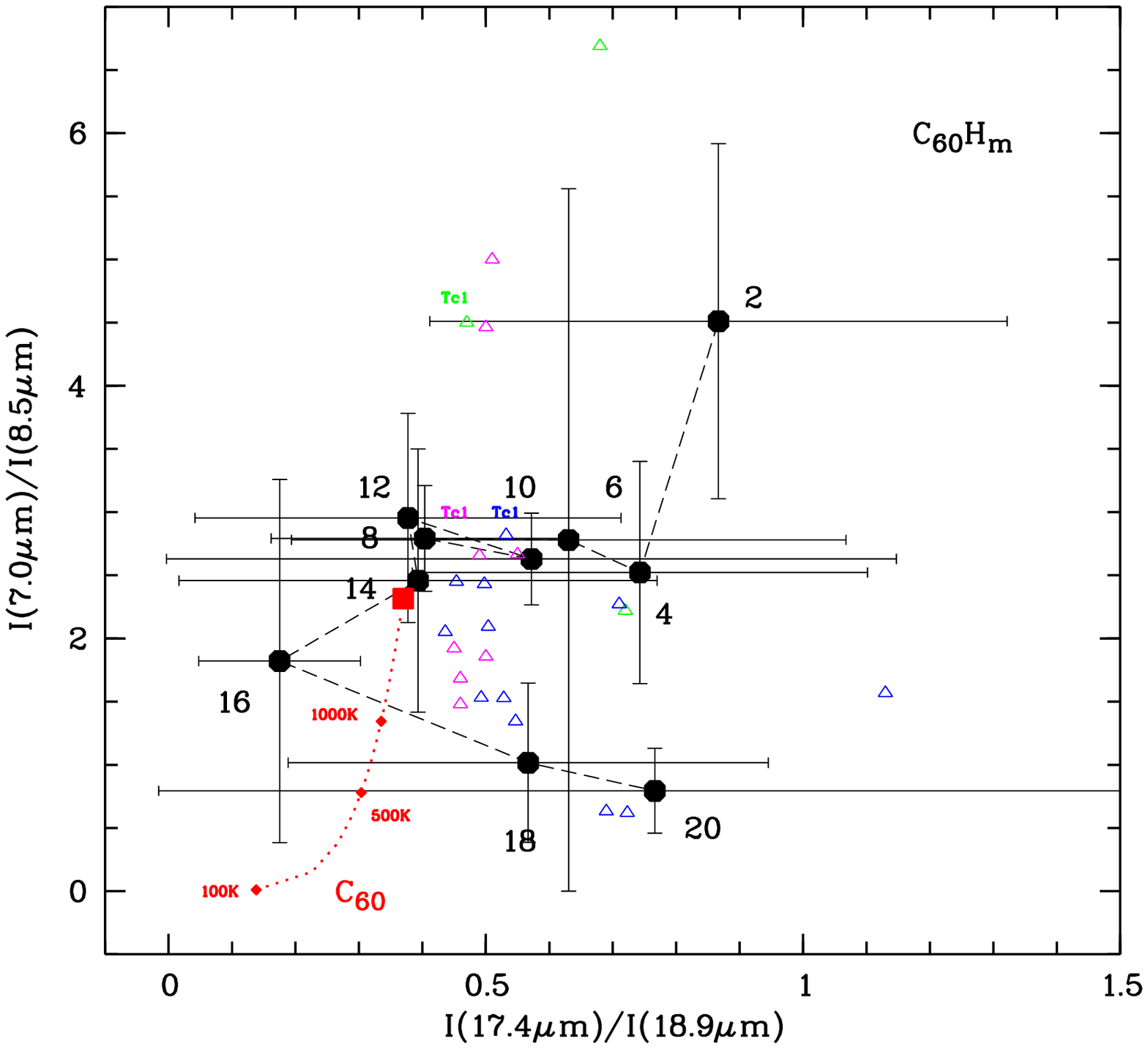, height=13cm}
\caption{
Intrinsic strength ratios of C$_{60}$H$_m$ (filled circles connected by dashed line), where the bars represent the standard deviations of the means of different isomers.  The marked numbers represent the $m$ value. The red square marks the position for C$_{60}$. The dotted curve represents the predicted ratios for a thermal excitation of C$_{60}$. The  green, blue, and magenta open triangles are the observed values taken from \citet{bc12}, \citet{gar12}, and \citet{ots14}, respectively. For the sake of clarity, the error bars of the observed values are not plotted.
}
\label{exc}
\end{figure*}

\begin{figure*}
\epsfig{file=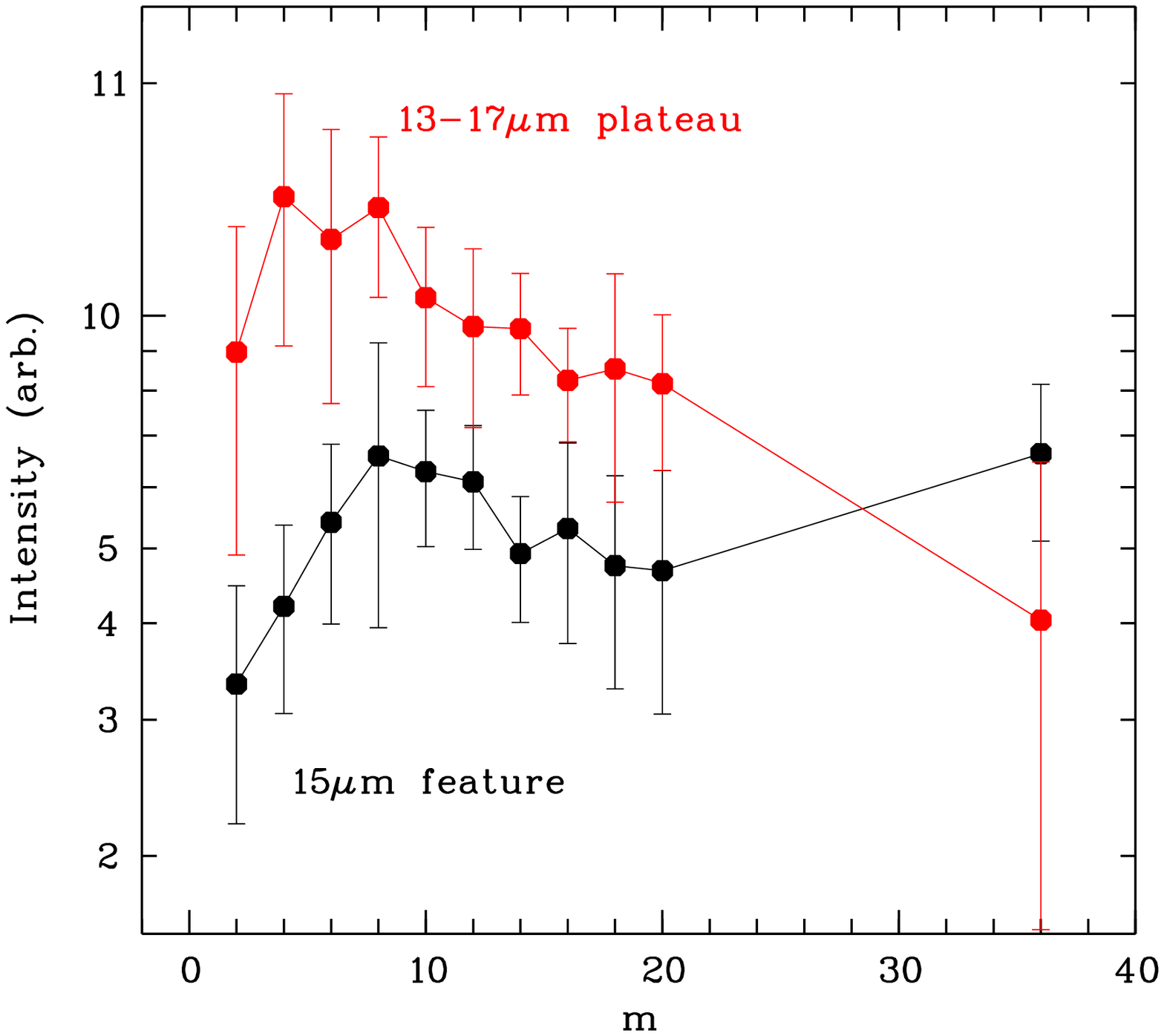, height=13cm}
\caption{The fluxes of the 15\,$\mu$m feature (black) and the 13--17\,$\mu$m plateau (red) versus  the $m$ values of  C$_{60}$H$_m$. The bars represent the standard deviations of the means of different isomers.
}
\label{15m}
\end{figure*}

\begin{figure*}
\epsfig{file=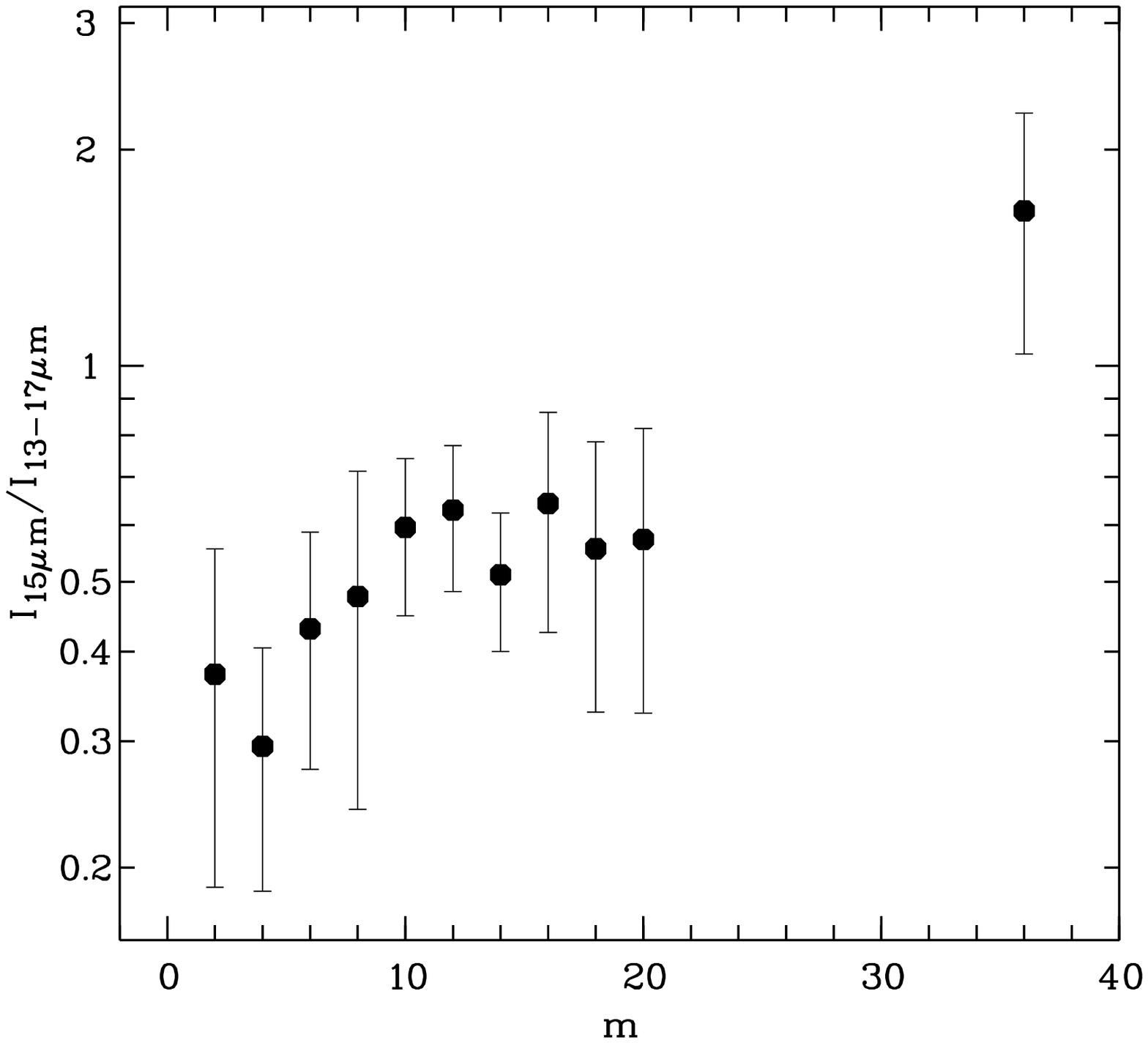, height=13cm}
\caption{The flux ratios between the 15\,$\mu$m feature and the 13--17\,$\mu$m plateau versus  the $m$ values of  C$_{60}$H$_m$. The bars represent the standard deviations of the means of different isomers.
}
\label{15m_pla}
\end{figure*}

\clearpage

\begin{figure*}
\begin{center}
\begin{tabular}{c}
\resizebox{140mm}{!}{\includegraphics{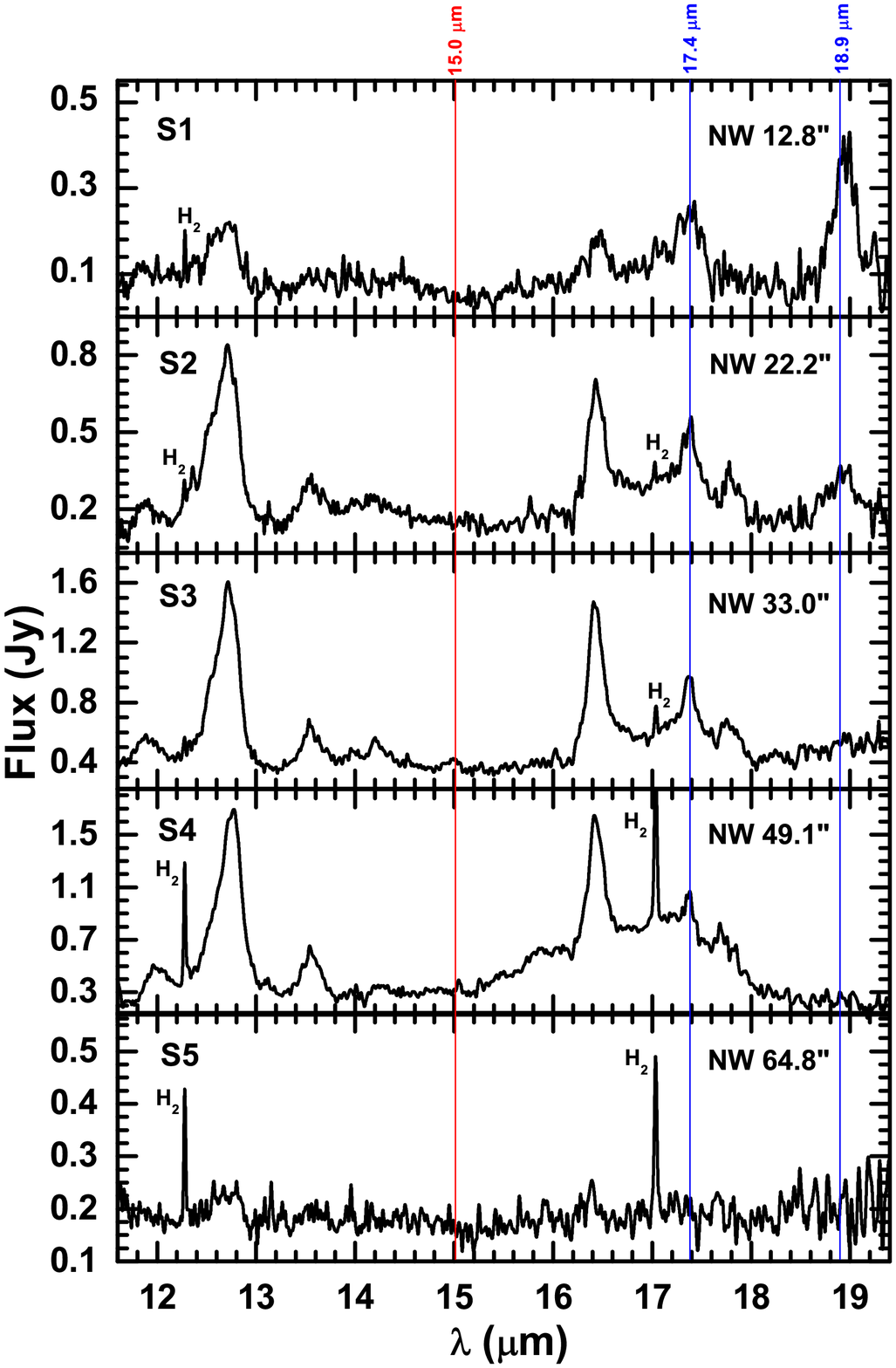}} \\
\end{tabular}
\end{center}
\caption{
Five {\it Spitzer}-IRS SH continuum-subtracted spectra for NGC 7023 at various distances from the central star.
The blue and red vertical lines indicate the positions of the C$_{60}$ bands and 15.0\,$\micron$ features, respectively. Note that the 17.4\,$\micron$ C$_{60}$ feature may be blended with an UIE band.  The panel for each spectrum S1--S5 is labeled with the slit offset to central source in the unit of arcseconds.  Two H$_{2}$ lines at 12.28 and 17.03\,$\micron$ are also marked.}
\label{ngc7023}
\end{figure*}

\begin{figure*}
\plotone{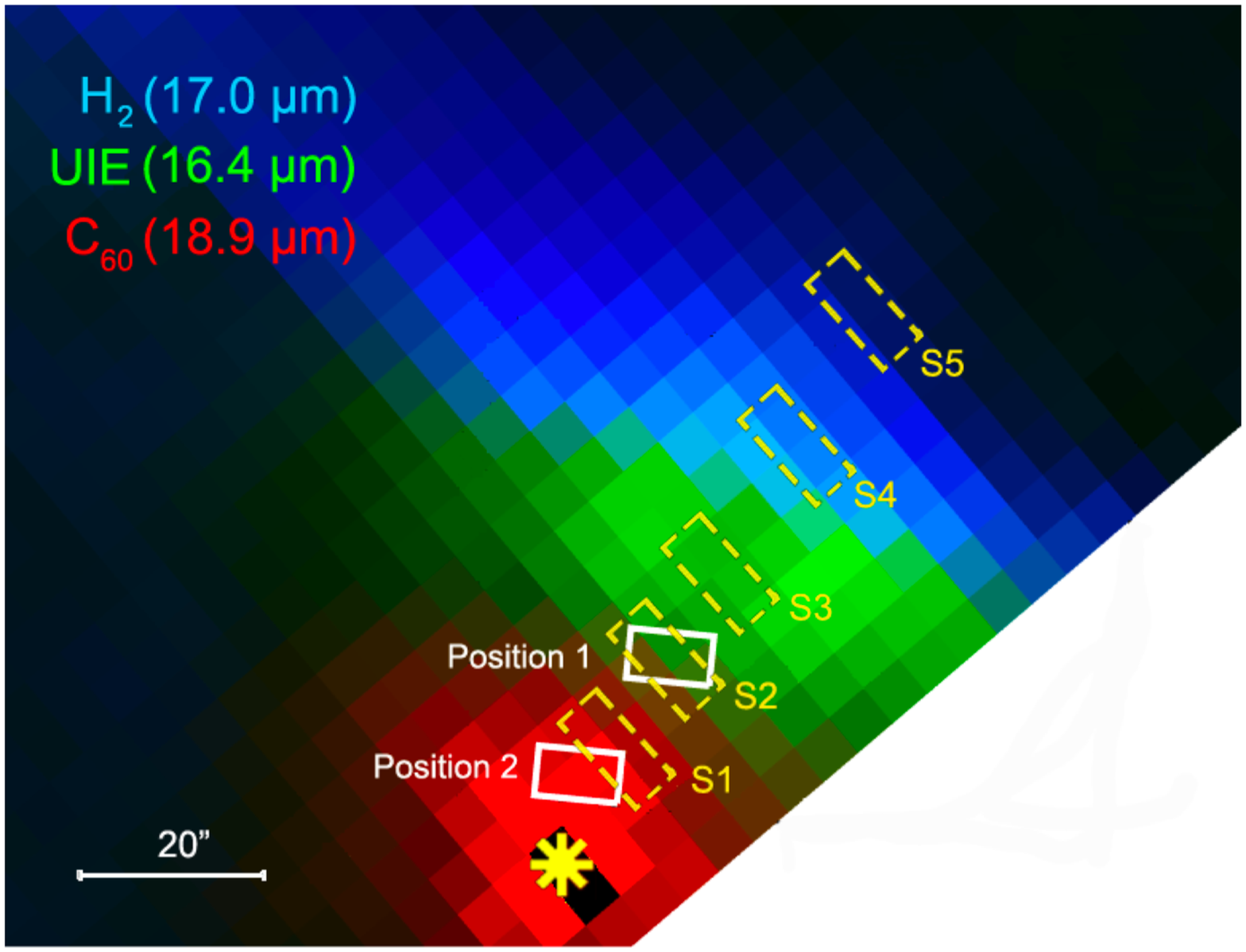}
\caption{{\it Spitzer} composite-color image of NGC 7023 nebula based on Fig. 1 in \citet{ber13}, constructed from components of {\it IRS} low-resolution (LL) module spectral cube. The image was mainly made from three integrated band intensities: H$_{2}$ line at 17.0 $\micron$ (shown in blue), the UIE band at 16.4 $\micron$ (green), and C$_{60}$ band at 18.9 $\micron$ (red). North is up and east is to the left. The white rectangles indicate the positions where \citet{ber13} extracted their spectra. The yellow dashed rectangles represent the regions where the IRS-SH spectra shown in Fig.~\ref{ngc7023} have been extracted. The position of the central star is also marked with yellow asterisk.
}
\label{map}
\end{figure*}

\begin{figure*}
\begin{center}
\begin{tabular}{c}
\resizebox{120mm}{!}{\includegraphics{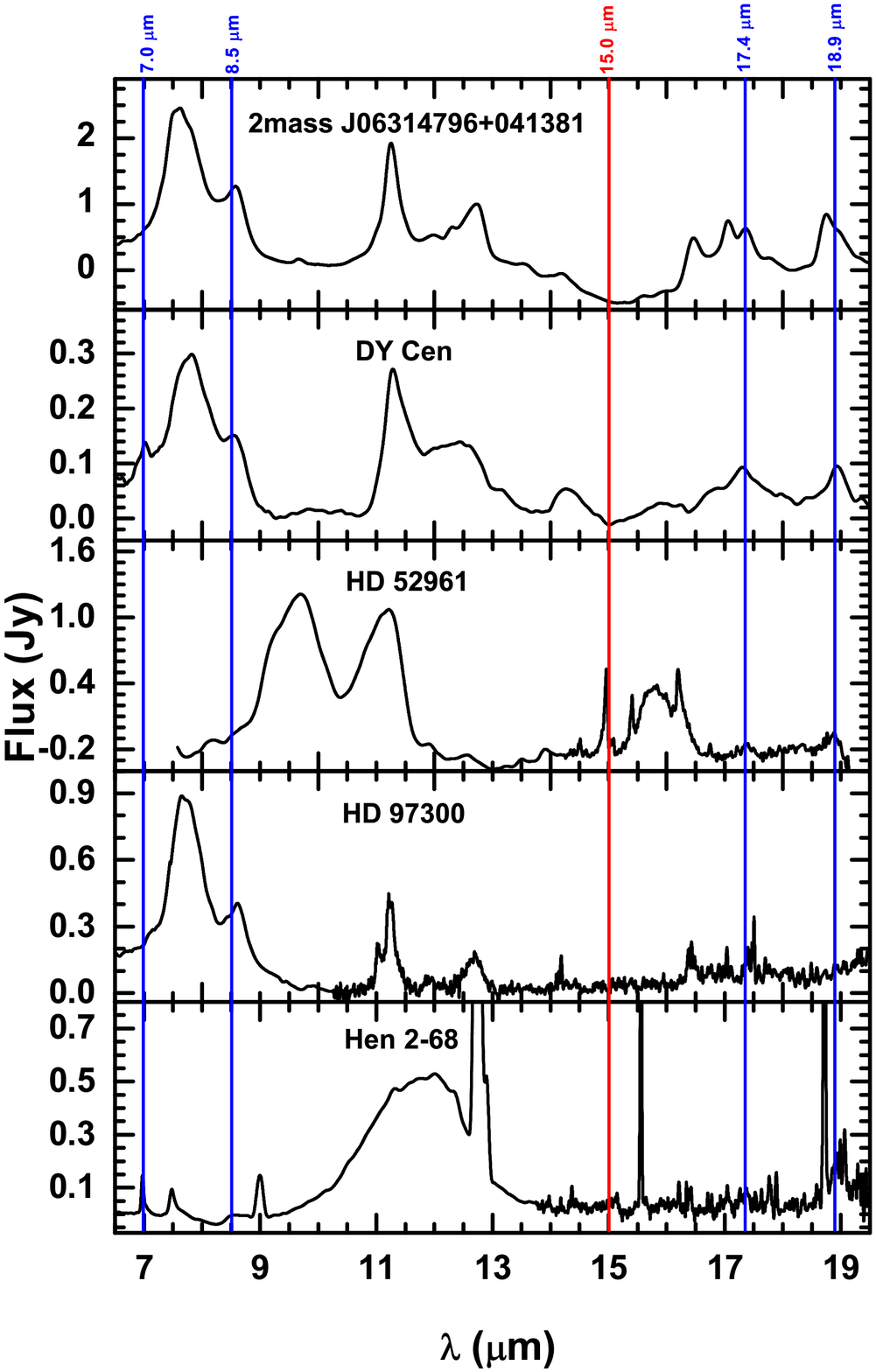}} \\
\end{tabular}
\end{center}
\caption{
Continuum-subtracted mid-infrared spectra  of known C$_{60}$ sources in the wavelength range of 6.5--19.5 $\micron$. The blue and red lines
indicate the positions of the C$_{60}$ bands and the 15\,$\mu$m feature, respectively. (Note that this figure has been abridged to fit the size limit of arXiv.
)}
\label{spectra}
\end{figure*}

%
%
%
%
%
%

\clearpage


\end{document}